%% file: main.tex
\RequirePackage{fix-cm}

\documentclass[smallextended]{svjour3}       

\smartqed  

\usepackage{graphicx}
\usepackage{amsmath}
\usepackage[caption=false]{subfig}
\usepackage{xcolor}
\usepackage{url}

\graphicspath{{./figures/}}

\begin{document}

\title{Talent Flow Analytics in Online Professional Network}




\author{Richard J. Oentaryo         \and
        Ee-Peng Lim \and
        Xavier Jayaraj Siddarth Ashok \and
        Philips Kokoh Prasetyo \and 
        Koon Han Ong$\dagger$ \and
        Zi Quan Lau$\dagger$ 
}


\institute{Richard J. Oentaryo \at
           McLaren Applied Technologies, Singapore \\
           \email{richard.oentaryo@mclaren.com}   
           \and
           Ee-Peng Lim, Xavier Jayaraj Siddarth Ashok, Philips Kokoh Prasetyo \at
           Living Analytics Research Centre, Singapore Management University \\
           \email{\{eplim,xaviera,pprasetyo\}@smu.edu.sg}
		   \and
           Koon Han Ong$\dagger$, Zi Quan Lau$\dagger$ \at
           Singapore Management University \\
           \email{koonhan.ong.2013@accountancy.smu.edu.sg} \\
           \email{ziquan.lau.2015@sis.smu.edu.sg} \\
           $\dagger$Work done during internship at Living Analytics Research Centre           
}

\date{Received: date / Accepted: date}

\maketitle

\begin{abstract}
Analyzing job hopping behavior is important for understanding job preference and career progression of working individuals. When analyzed at the workforce population level, job hop analysis helps to gain insights of talent flow among different jobs and organizations. Traditionally, surveys are conducted on job seekers and employers to study job hop behavior. Beyond surveys, job hop behavior can also be studied in a highly scalable and timely manner using a data driven approach in response to fast-changing job landscape. Fortunately, the advent of online professional networks (OPNs) has made it possible to perform a large-scale analysis of talent flow. In this paper, we present a new data analytics framework to analyze the talent flow patterns of close to 1 million working professionals from three different countries/regions using their publicly-accessible profiles in an established OPN.  As OPN data are originally generated for professional networking applications, our proposed framework re-purposes the same data for a different analytics task. Prior to performing job hop analysis, we devise a job title normalization procedure to mitigate the amount of noise in the OPN data. We then devise several metrics to measure the amount of work experience required to take up a job, to determine that the duration of a job's existence (also known as the job age), and the correlation between the above metric and propensity of hopping. We also study how job hop behavior is related to job promotion/demotion. Lastly, we perform connectivity analysis at job and organization levels to derive insights on talent flow as well as job and organizational competitiveness.

\end{abstract}

\input{introduction}
\input{related_work}
\input{framework}
\input{network}
\input{method}
\input{insights}
\input{conclusion}

\begin{acknowledgements}
This research work is supported by the National Research Foundation, Prime Minister's Office, Singapore under its International Research Centres in Singapore Funding Initiative.
\end{acknowledgements}

\bibliographystyle{spmpsci}      
\bibliography{references}   

\end{document}

%% file: introduction.tex
\section{Introduction}
\label{introduction}

Job hop is a common behavior observed in any workforce. As a person hops from jobs to jobs, he or she acquires new skills and potentially gains higher income.  Every job hop captures an important decision made by the person as well as an attempt of the hiring organization to acquire talent. When job hop behavior is analyzed at the workforce level, it will yield insights about the workforce, job pool and employers.

Such insights have been traditionally obtained using surveys on employers and job seekers. For example, the US Bureau of Labor Statistics (BLS) conducts annual surveys with approximately 146,000 businesses and government agencies to collect employment data\footnote{\scriptsize \url{www.bls.gov/ces/}}.  The surveys yield useful information about job demand, job supply, income, working hours, etc..  While surveys can be a powerful instrument to gather direct user input, they are usually not scalable.  In the case of the BLS surveys, they cover less than $1\%$ of all U.S businesses. Moreover, as fast-changing technologies (such as sharing economy \cite{Hamari2015}) begin to impact job demand quickly, it is critical to explore new ways to obtain job related insights.

Past studies \cite{Joseph:2012,Maier:2015,Huayu:KDD2017} also tend to study jobs and organizations in isolation, without considering them as connected networks and how the networks capture talent flows from jobs to other jobs, and from organizations to other organizations. A lack of this network view prevents us from analyzing the ways people build their career, and competition among organizations for talent. For example, some job changes could be promotions, while others could just be lateral and even demotions. The network view is also crucial in studying how the competitions among jobs and organizations would eventually impact job creation and talent attraction.

In contrast, online professional networks (OPNs) are fast becoming a marketplace for resume posting, candidate hunting, and job searching. Representative examples of OPN are LinkedIn, Xing and Viadeo\footnote{\scriptsize \textbf{LinkedIn} -- \url{www.linkedin.com}; \textbf{Xing} -- \url{www.xing.com}; \textbf{Viadeo} -- \url{www.viadeo.com}}. Detailed job activity data at the individual user level are now publicly available in these OPNs, as soon as the users update their profiles. These data can be analyzed to derive interesting behavioral insights about jobs and organizations, as well as to build services that can benefit both employers and job seekers, e.g., a service that helps employers find suitable employees and another service that helps job seekers find suitable jobs.

\textbf{Objectives}.
In this work, we focus on using data from one of the world's largest OPNs to analyze job hops and talent flow. To support our analysis on hops within an organization and across organizations, we first develop several metrics that measure the amount of experience required for a job and how established/recent a job is, from the viewpoint of the people holding the job.

We also aim at studying how the job hop behavior of a workforce is related to job promotion/demotion. This is a topic often discussed based on anecdotal examples \cite{Alper:1994,Hamori:2010}. A better approach is to conduct a large-scale data science study. This will give much broader insights on job hop patterns particularly useful in human resource recruitment and career coaching.

Finally, our research aims at analyzing talent flow based on job hop behavior and measuring the capabilities of each job and organization in attracting, supplying, and competing for human capital. To this end, we create a weighted directed hop network among jobs and organizations, develop different centrality measures for the job and organization nodes, and evaluate them by manual inspection or by comparing with other attributes such as organization size.

\textbf{Contributions}. To accomplish the above objectives, we develop a new data analytics framework to clean, aggregate, and derive talent flow insights from OPN data. The proposed framework constitutes a generalization of our earlier work~\cite{Oentaryo:ICDMW2017}, featuring two major extensions:
\begin{itemize}
\item Prior to talent flow analytics, we introduce job title translation, parsing, and normalization steps as additional data cleaning/pre-processing steps in order to mitigate noise in our OPN datasets.
\item To demonstrate the applicability of our proposed framework, we conduct an extended study using OPN datasets of working professionals from three countries/regions (i.e., Singapore, Switzerland and Hong Kong) with diverse workforce profiles. Our study reveals a number of interesting findings and insights that unveil similarities and differences among countries/regions.
\end{itemize}
All in all, the main contributions of this work are as follows:
\begin{itemize}
\item We present a talent flow analytics framework to facilitate a data science approach to analyze talent flow among jobs and organizations. This outlines the essential steps to re-purpose the OPN data for talent flow study that complements the traditional surveys.

\item We devise several key metrics to analyze talent flow networks, with the aim to answer several research questions connecting talent flow with career progression, user attributes (e.g., working experience), and user career behavior (e.g., promotion and demotion). We conduct empirical studies using the key metrics to explain some patterns in our datasets.

\item We justify the applicability of our approach through extensive empirical study using OPN datasets from three different countries/regions. The results reveal interesting insights on the similarities and differences of the talent flow patterns across countries/regions. 
\end{itemize}

\textbf{Paper outline}. The remainder of this paper is organized as follows. Section \ref{sec:related} first provides a survey of related works. Section \ref{sec:overview} in turn describes the proposed talent flow analytics framework and the dataset used in our study. Details of the talent flow network construction and talent analytics approaches are described in Sections \ref{sec:network_construction} and \ref{sec:analytics_methods} respectively. Section \ref{sec:insights} presents the key insights and discussion on the results. Finally, Section \ref{sec:conclusion} concludes this paper.

%% file: related_work.tex
\section{Related Work}
\label{sec:related}

Research on job and workforce movements has been around for decades \cite{NBERw2649,LabourMobility,moscarini2007occupational,fuller2008job,Joseph:2012}.  Topel \emph{et al.} \cite{NBERw2649} analyzed $15$ years of job changing and wage growth of young men from Longitudinal Employee-Employer data. Long \emph{et al.}~\cite{LabourMobility} studied the labor mobility in Europe and the U.S. Moscarini \emph{et al.} \cite{moscarini2007occupational} measured worker mobility across occupations and jobs in the monthly Current Population Survey data from 1979 to 2006. More recent survey-based studies \cite{fuller2008job,Joseph:2012,Schawbel:2013} have revealed that the younger employees are more likely to switch jobs and employers/companies than the older ones. Friedell \emph{et al.}~\cite{friedell2011} showed that there is a discrepancy between younger generations' expectations in the workplace and older generations' perception of those expectations.

In general, these studies traditionally relied on surveys, census, and other data such as tax lists and population registers, which require extensive and time-consuming efforts to collect. Often times, surveys may focus on selected workforce segments or industries. Such an approach thus cannot be easily scaled up or replicated across many segments/industries.

With the wide adoption of OPNs, there is a rapidly-growing interest to mine the online user data from the OPNs to understand job and workforce movements as well as career growth. For example, State \emph{et al.} \cite{State:Socinfo2014} analyzed the migration trends of professional workers into the U.S. Xu \emph{et al.} \cite{Xu:ICDM2015} combined work experiences from OPNs and check-in records from location-based social networks to predict job change occasions. Chaudhury \emph{et al.} \cite{Chaudhury:WWW2016} analyzed the growth patterns of the ego-network of new employees in companies.

An important aspect in OPNs is job hop. Job hop data capture a wide range of signals that can help understand the performances of organizations, talent sources, job market, professional profiles, as well as career advancement. Cheng \emph{et al.} \cite{Cheng:KDD2013} modeled job hop activities to rank influential companies. Xu \emph{et al.} \cite{Xu:KDD2016} generated and analyzed job hop networks to identify talent circles. Kapur \emph{et al.} \cite{Kapur:KDD2016} devised a talent flow graph to rank universities based on the career outcomes of their graduates. They applied their approach to two specific workforce segments: investment banker and software developer. 

Users' career paths have also been utilized to model professional similarity for use in job recruitment process \cite{Xu:KDD2014}. In this work, a sequence alignment method was used to quantify similarity between two career paths. Liu \emph{et al.} \cite{Liu:AAAI2016} devised a multi-source learning framework that combines information from multiple social networks to predict the career path of a user. While the approach is interesting, their work focused only on four job categories, namely software engineer, sales, consultant, and marketing. Recently, Li \emph{et al.}~\cite{Huayu:KDD2017} proposed a survival analysis approach to model career paths for turnover and career progression. However, their study was conducted on within-organization career paths (i.e., inside a company) for talent management.

Additionally, career trajectory similarity has been proposed to identify individuals who share similar career histories with some given user provided ideal candidates so that the former can be returned as talent search results \cite{Xu:KDD2014,HaThuc:WWW2016}. Xu et al. \cite{Xu:KDD2016} defined a job transition network with vertices and edges representing organization and talent flow between two organizations for a time period respectively. From the network, talent circles each covering a set of organizations with similar talent exchange patterns are detected.  It has been shown that talent circles can improve talent recruitment and job search.

On a related track, Xu et al. \cite{Xu:ICDM2015} analyzed job change patterns using OPN data and correlated these patterns with human activity data from a location-based social networking site. They also proposed a set of features to predict future job changes to be made by some employee.

\textbf{Our research}. The work presented in this paper differs from the above-mentioned works in several unique ways. Firstly, we introduce quantitative metrics to measure how much work experience is required to take up a job and how recent/established a job is, and examine their relationships with the propensity of hopping. Secondly, we compute the level gain of job hops so as to analyze promotion/demotion of employees which, to our best knowledge, has been missing in the previous studies. Additionally, we perform an extensive study on talent flow and competition by analyzing both job-level and organization-level hop networks, without being restricted to specific workforce segments or industries. Last but not least, our study involves multiple countries or regions, thus providing more comprehensive insights on how talent flows compare among different countries or regions.

%% file: framework.tex
\section{Dataset and Method}
\label{sec:overview}

This section provides an overview of the OPN dataset considered in this work as well as of our proposed approach for talent flow analytics.

\subsection{Dataset}
\label{sub:data}

\begin{table}[!t]
\caption{Statistics of the OPN datasets used in this study}
\label{tab:basic_stats}
\begin{tabular}{|l|r|r|r|}
\hline
Statistics & Singapore & Switzerland & Hong Kong \\
\hline
Number of user profiles 		& 1,674,432 & 1,419,129 & 432,525\\
Number of core user profiles & 502,123 & 377,590 & 82,672\\
Number of organizations		& 151,638 & 124,462 & 45,652\\ 

\hline
\end{tabular}
\end{table}

To facilitate our empirical studies, we extract online public profiles from one of the world's largest OPNs. In particular, our data collection involves extracting the public profiles of all OPN users in three countries/regions, i.e., Singapore, Switzerland, and Hong Kong. We collect a list of public profiles from public users directory associated with the target city or region. We also extract organization profiles mentioned in these public user profiles.  As shown in Table~\ref{tab:basic_stats}, the datasets consist of 1.6M and 1.4M user profiles involving about 151K and 124K organizations for Singapore and Switzerland respectively. The Hong Kong dataset is smaller with 432K user profiles and 45K organizations. 

Meanwhile, the core users in Table~\ref{tab:basic_stats} refer to users who have at least one entry in the education, experience, and skill fields. As per Table~\ref{tab:basic_stats}, we have 502K core users in Singapore, 377K core users in Switzerland, and 82K core users in Hong Kong. For some analyses that we perform in this work, such as on work experience and job level (see Section~\ref{sub:jobattribute-analysis}), these fields need to be available. In these cases, the relevant metrics are computed based on the core users only, and non-core users are excluded. For all other analyses such as hop extraction, job title normalization, and talent flow network construction (cf. Section 3.2), we include all users that are found in our  data.

\subsection{Talent Flow Analytics Framework}
\label{sec:framework}

\begin{figure}[!t]
  \includegraphics[width=1.0\columnwidth]{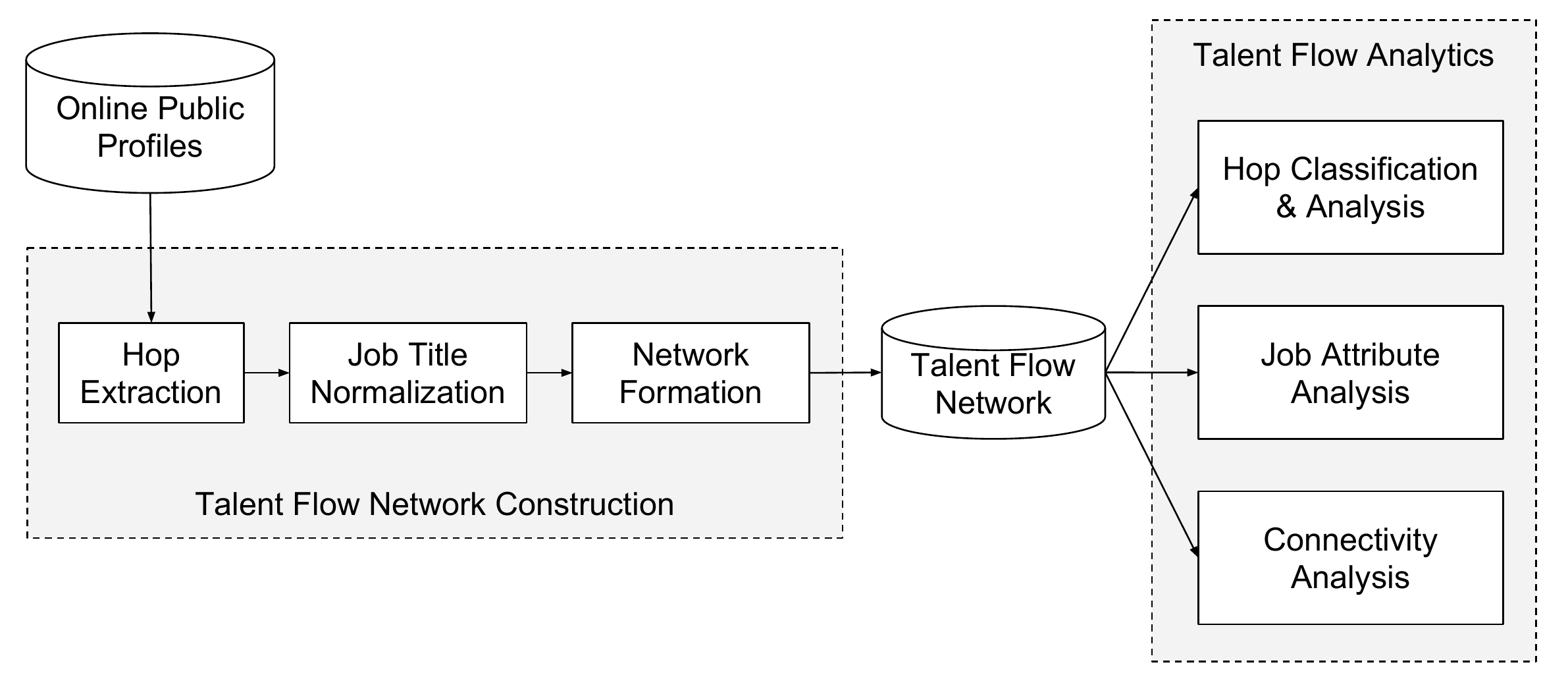}
  \centering
  \caption{Overview of the proposed talent flow analytics framework}
  \label{fig:analytics-overview}
\end{figure}

Our proposed talent flow analytics framework, as depicted in Figure~\ref{fig:analytics-overview}, consists of two key phases: \emph{talent flow network construction} and \emph{talent flow analytics}. In the first phase, we construct a talent flow network from the online public profiles. This comprises three steps: \emph{hop extraction}, \emph{job title normalization}, and \emph{network formation}. In the hop extraction step, we collect all job hops from online public profiles. During job title normalization, we reduce the duplicate job titles which occur due to variations in language, small typos, and non standardized writing. For instance, ``finance manager'' is the same as ``manager, finance'', ``manager - finance'', ``finance mananger'', and ``finance manger''. (Note that the last two finance manager variations are due to typos.) 

After job title normalization, we represent each job by its normalized job title and the company. We then craft the talent flow network based on transitions between jobs and perform talent flow analytics that encompasses three types of analysis, namely hop classification and analysis, job attribute analysis, and connectivity analysis. Each analysis studies a specific aspect of talent flow network. Firstly, hop classification and analysis focuses on analyzing types of job hop activity in the network whether the hop is an internal or external hop. Hop analysis will reveal the pattern of internal and external hops.  Secondly, job attribute analysis aims to analyze talent flow with respect to job attributes and job hop attributes such as promotion and demotion. Finally, connectivity analysis strives to analyze talent flow behavior at the network level allowing us to determine important jobs and organizations. We describe the talent flow network construction and analytics phases in greater detail in Sections \ref{sec:network_construction} and \ref{sec:analytics_methods} respectively.

%% file: network.tex
\section{Talent Flow Network Construction}
\label{sec:network_construction}

In this section, we describe in greater detail the steps involved in the construction of a talent flow network (as per Fig. \ref{fig:analytics-overview}).

\subsection{Hop Extraction}
\label{sub:hop-extraction}

The first step to construct the talent flow network is to extract \emph{job hop} from the online public profiles. Each \emph{job} in talent flow network is defined as a triplet $(t, c, i)$, which represents job title $t$ at organization $c$ in industry $i$ respectively. Note that each organization $c$ belongs to a unique industry $i$. We then define a \emph{job hop} as a transition from one job to another job with \emph{non-overlapping} time period. A job hop represents a talent flow from one job (or organization) to another job (or organization), and a collection of job hops forms a talent flow network (see Section~\ref{sub:network-construction}). 

\begin{figure}[!t]
\centering
\includegraphics[width=1.0\columnwidth]{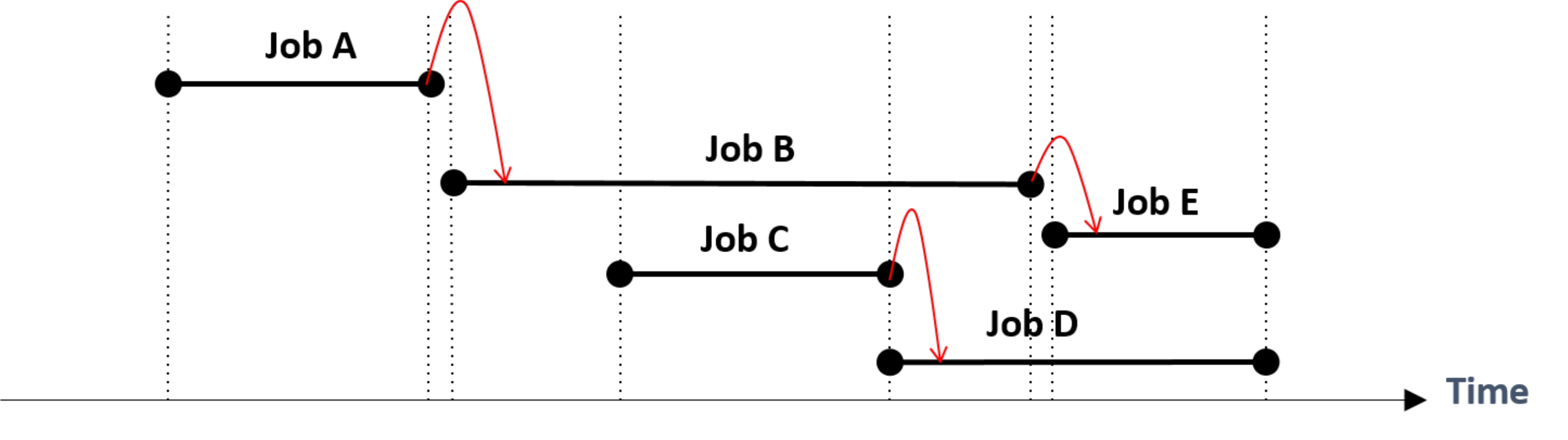}
\caption{Definition of job hop}
\label{fig:hop_definition}
\end{figure}

Fig. \ref{fig:hop_definition} shows an example of an OPN user who lists five jobs $A$, $B$, $C$, $D$ and $E$ in his profile. In this case, the user is regarded as having only three hops, i.e., from job $A$ to job $B$, from $B$ to $E$, and from $C$ to $D$. There is no hop from $B$ to $C$, or from $B$ to $D$, or from $D$ to $E$, since they take place in an overlapping time period and are likely to be side activities of the user. To capture as many distinct job titles as possible for the next step (i.e., job title normalization), we include all users for this step, not only the core users.

Table~\ref{tab:hop_stats} summarizes the statistics of the extracted job hops showing the number of job hops, and distinct job titles\footnote{Distinct job titles refer to unique string of job titles. For example, ``Java Developer'' and ``Java Software Engineer'' are considered as different job titles, even though semantically they might correspond to the same job.} for Singapore, Switzerland, and Hong Kong datasets. From the extracted job hops, we collect more than 795,000 distinct job titles in Singapore, 1 million distinct job titles in Switzerland, and 195,000 distinct job titles in Hong Kong. We discovered that job titles in Switzerland consist of many languages such as English, French, German, Italian, Spanish and Portuguese. We also found that the majority of job titles in Hong Kong are in English and Mandarin. 

\begin{table}[!t]
\caption{Statistics of the extracted job hops}
\label{tab:hop_stats}
\begin{tabular}{|l|r|r|r|}
\hline
Statistics & Singapore & Switzerland & Hong Kong \\
\hline
Number of hops 	& 4,561,881 & 4,027,083 & 917,617\\
Number of distinct job titles & 795,249 & 1,094,061 & 195,659\\
Number of distinct job titles (min\_sup $\geq$ 10) & 38,966 & 38,220 & 8,982 \\
\hline
\end{tabular}
\end{table}

Although we have a large numbers of distinct job titles, the job titles are typically very noisy, owing to two main reasons:
\begin{itemize}
\item \textbf{Data sparsity}. This could be due to either poor/inaccurate naming of job titles or less popular jobs that have small number of occurrences. The first issue can be resolved by job title parsing and normalization, which we will describe in Section \ref{sub:title-normalization}. In contrast, the latter issue cannot be resolved by job title normalization. To mitigate this effect, we define a threshold for each job title in the extracted hop collections to be at least 10 instances.

\item \textbf{Job title variation.} This variation is typically caused by language diversity and non-standardized job title naming. This can be addressed by job title parsing and normalization as well as job title translation for non-English job titles. This is further elaborated in Section \ref{sub:title-normalization}.
\end{itemize}

\subsection{Job Title Normalization}
\label{sub:title-normalization}

We develop a parser to normalize job title on the extracted hop collection. Job title normalization is important to reduce variations of same job title such as 'research director' and 'director of research'. For a given job title, the parser normalizes the title into its constituent parts, allowing us to extract important \emph{functional} components inside a given job title. We define the constituent parts of a given job title as follows:

\begin{enumerate}
\item \textbf{Primary function}: indicates the main job role. Each job title must have at least one primary function. Thus, this part is compulsory.
\item \textbf{Domain}: indicates the domain of a job role. This part is optional.
\item \textbf{Position}: indicates the seniority level of a job role. This part is optional.
\item \textbf{Secondary function}: indicates secondary job role. This part is optional.
\item \textbf{Additional information}: indicates extra information about the job title. It commonly appears inside a bracket. This part is optional.
\end{enumerate}

\begin{figure}[!t]
  \centering
  \includegraphics[width=0.8\columnwidth]{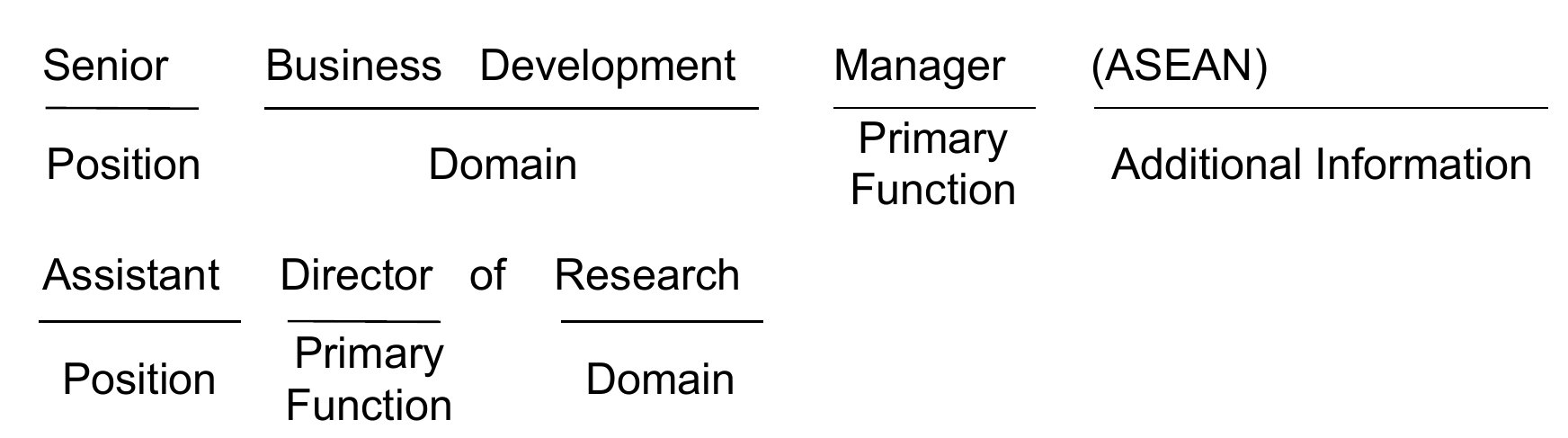}
  \caption{Examples of parsed job titles}
  \label{fig:parsedjobtitle}
\end{figure}

To build such a parser, we devise grammar rules to cover valid job title syntax. The entire parsing process can be broken down into several steps:
\begin{enumerate}
\item \textbf{Lexical analysis}: In this step, a lexical analyzer (lexer) tokenizes a job title input using defined regular expressions and matches them against dictionary files. 
\item \textbf{Syntax tree generation}: Using the tokens from lexer, the parser subsequently checks if these tokens adhere to the grammar rules and validates the syntax. If the syntax is valid, the concrete syntax tree will be generated from the selected rules.
\item \textbf{Extraction}: In this step, the constituent parts of the job title are extracted from the concrete syntax tree. 
\end{enumerate}
We implement the parser using PLY\footnote{http://www.dabeaz.com/ply/} parser tool. Few examples of the parsed job titles are shown in Fig.~\ref{fig:parsedjobtitle}. Job titles that fail the lexical analysis and syntax tree generation steps are considered to have parsing errors.

Before parsing the job titles, some efforts are required to handle the non-English job titles in Switzerland and Hong Kong datasets. We use the Google Translate API\footnote{We use the googletrans Python wrapper for Google Translate API - \url{https://pypi.python.org/pypi/googletrans}} to translate non-English job titles to English. The translation results are then validated by our job title parser. In this case, we expect the valid translated job titles to be parseable. 

\begin{figure}[!t]
\centering
\subfloat[Parsing error rates for Singapore data\label{fig:sg_parser_errors}]{
  \includegraphics[width=0.45\columnwidth]{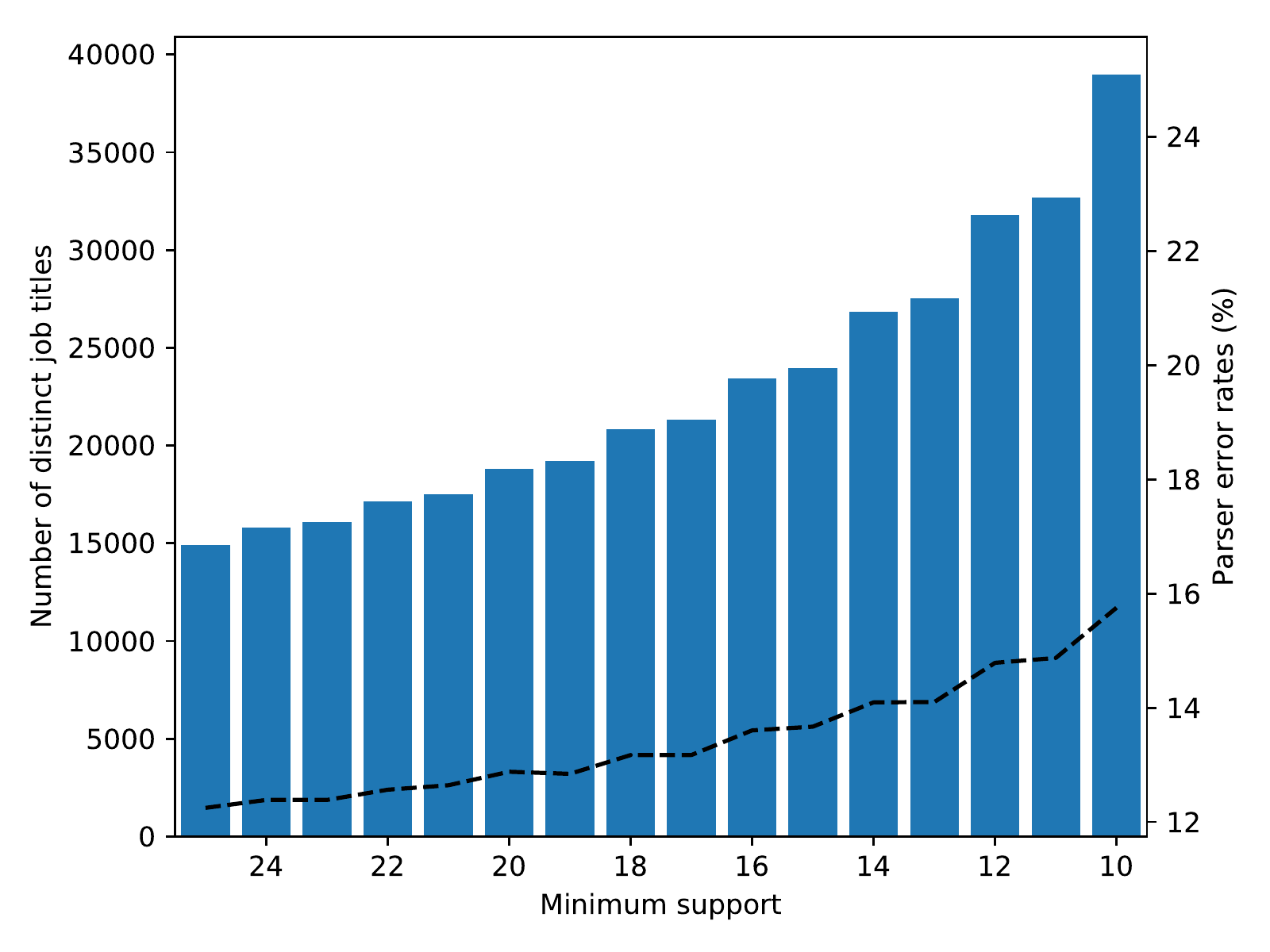} 
}\hfill
\subfloat[Parsed job titles for Singapore data\label{fig:sg_num_parsed}]{
  \includegraphics[width=0.45\columnwidth]{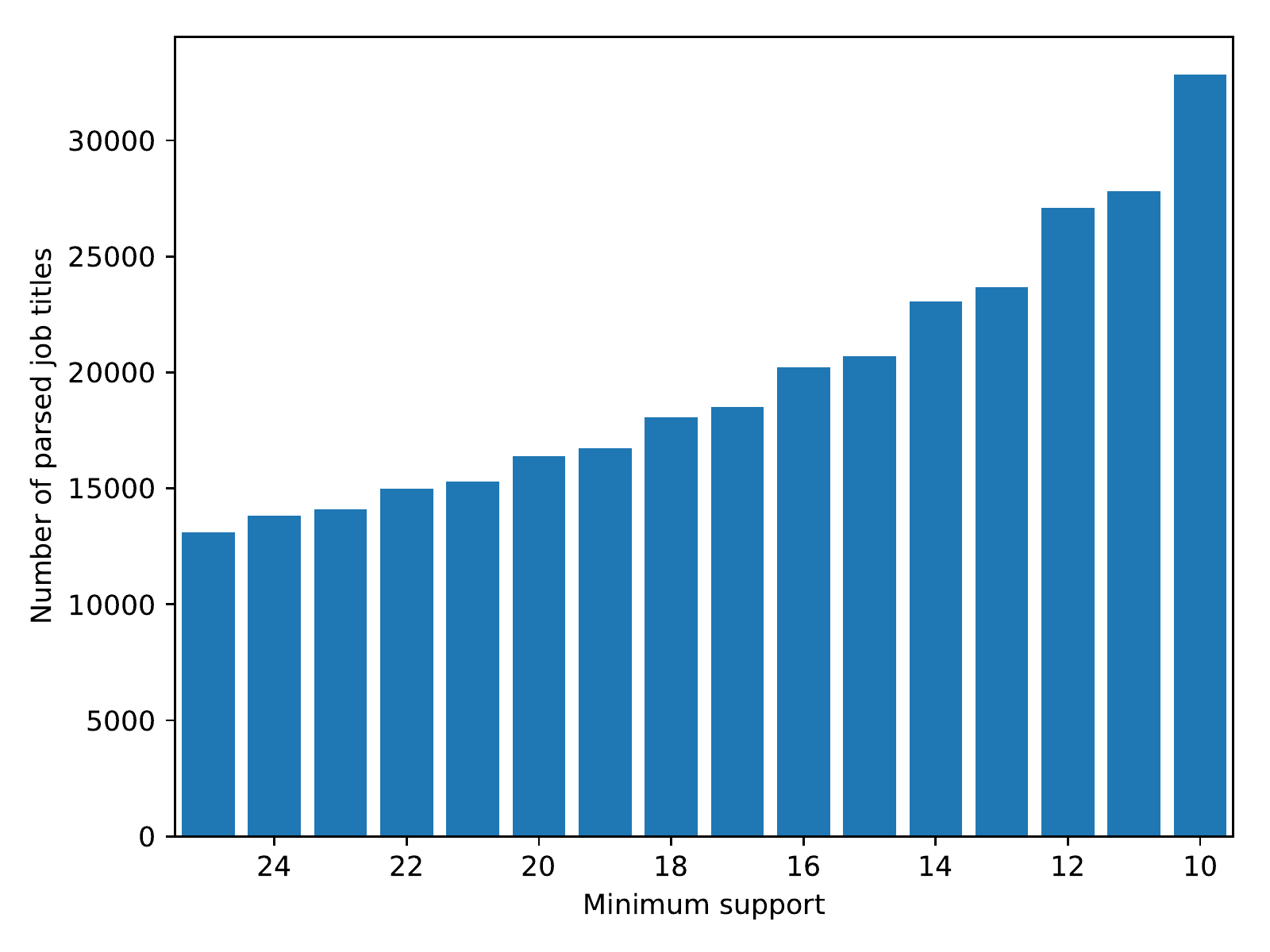} 
}\vfill
\subfloat[Parsing error rates for Switzerland data\label{fig:ch_parser_errors}]{
  \includegraphics[width=0.45\columnwidth]{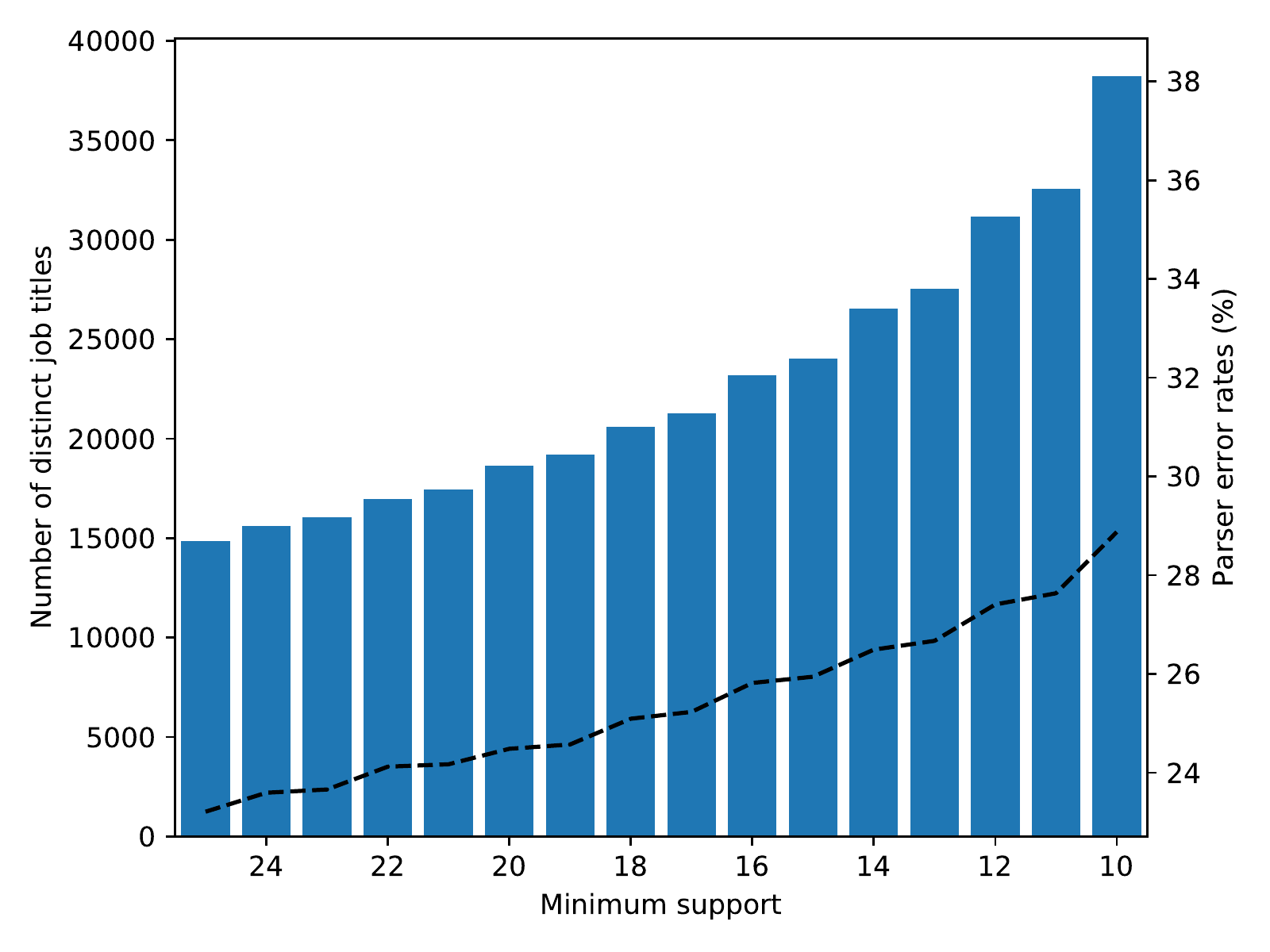}
}\hfill
\subfloat[Parsed job titles for Switzerland data\label{fig:ch_num_parsed}]{
  \includegraphics[width=0.45\columnwidth]{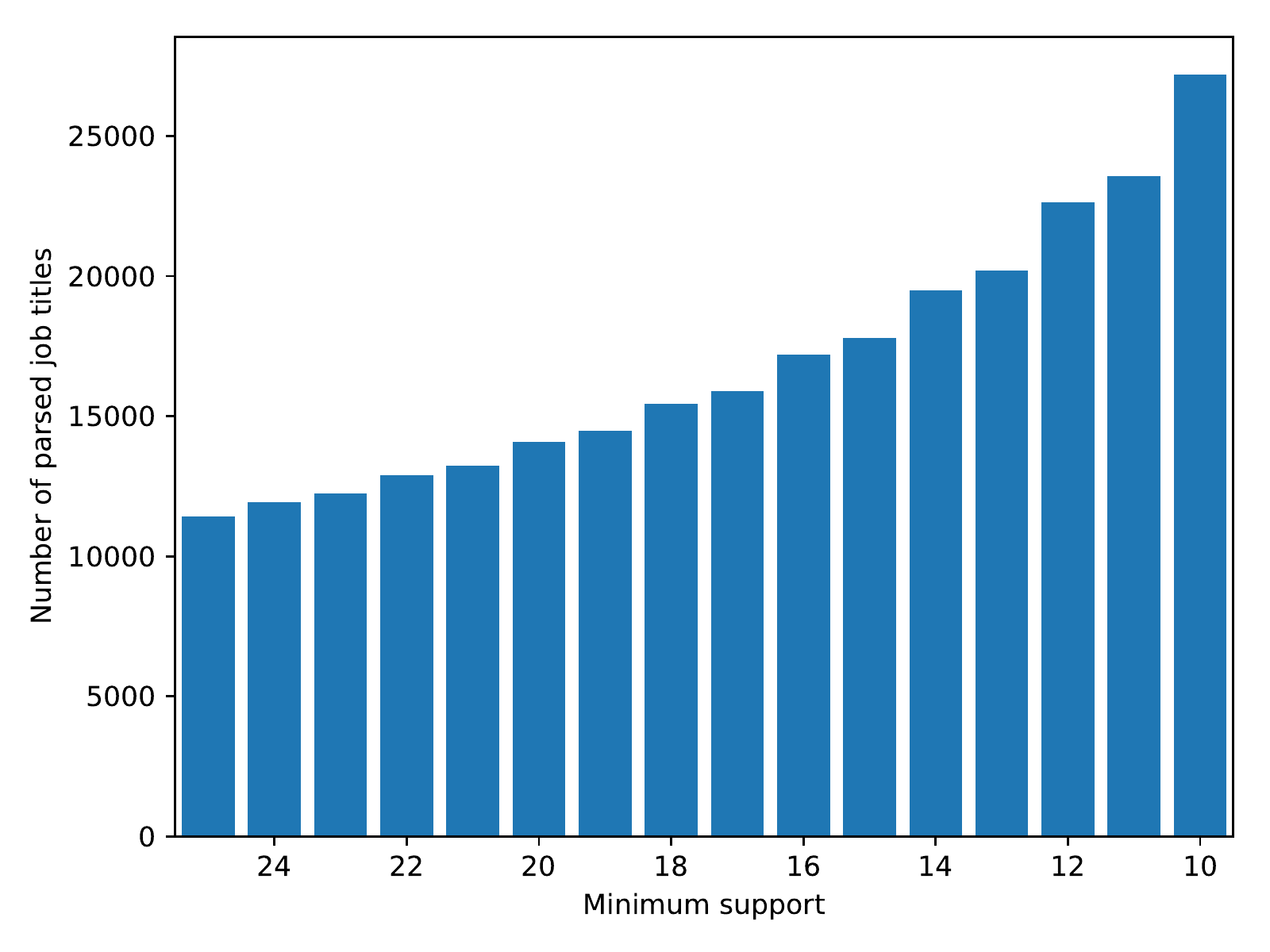} 
}\vfill
\subfloat[Parsing error rates for Hong Kong data\label{fig:hk_parser_errors}]{
  \includegraphics[width=0.45\columnwidth]{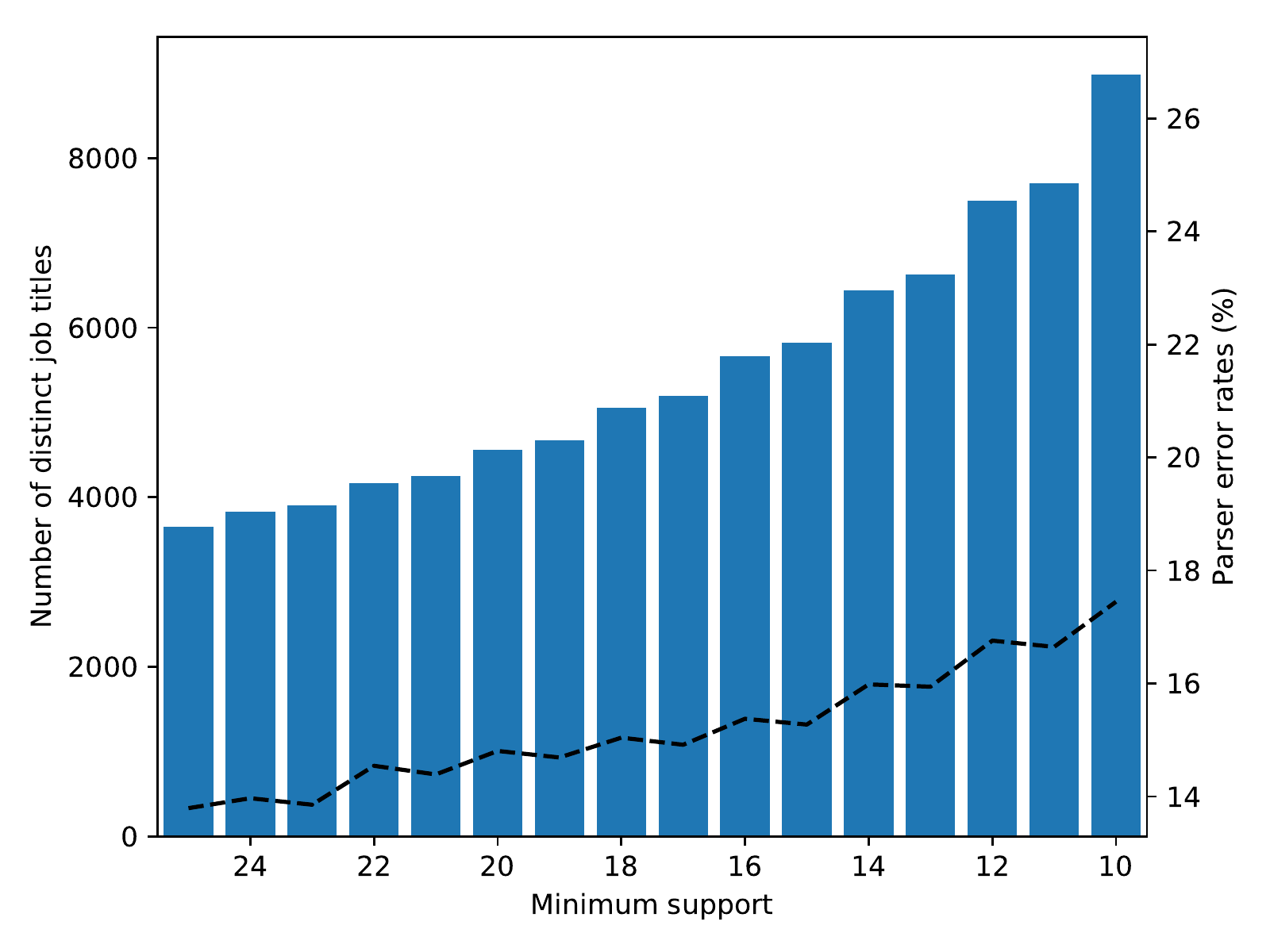}
}\hfill
\subfloat[Parsed job titles for Hong Kong data\label{fig:hk_num_parsed}]{
  \includegraphics[width=0.45\columnwidth]{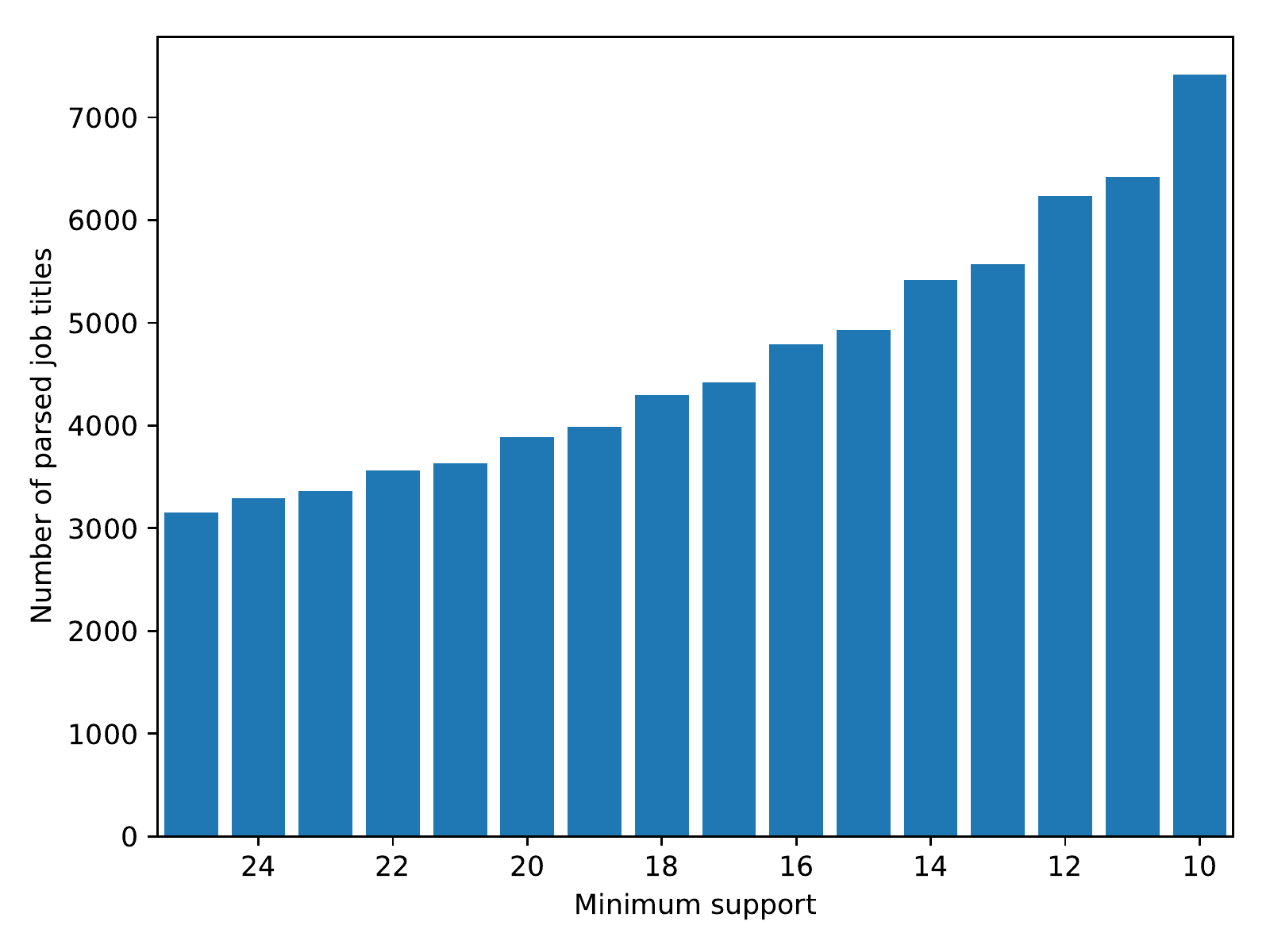} 
}
\caption{Results of job title parsing}
\label{fig:parser_errors}
\end{figure}

Fig.~\ref{fig:parser_errors} depicts the parsing error rates and the number of parsed job titles at different job title minimum support values on each dataset. In particular, Fig.~\ref{fig:parser_errors}(a), (c) and (e) show the number of distinct job titles and job parsing error rates at different minimum supports, while Fig.~\ref{fig:parser_errors}(b), (d) and (f) show the corresponding numbers of parsed job titles. We can see that, the lower the minimum support, the higher the number of distinct job titles, but the parsing error rates increase. At the minimum support of 10, the parsing error reaches 15.75\% for Singapore, 28.87\% for Switzerland, and 17.44\% for Hong Kong datasets. Following these processes, we obtain 32,828 parsed job titles in Singapore, 27,184 in Switzerland, and 7,415  in Hong Kong.

With job title parsing, it is expected that different job titles with the same constituent parts would map to the same parsed result. Among these job titles, we pick the most popular one as the normalized (i.e., canonical) job title, and use it to substitute all the other job titles with the same constituent parts. Table~\ref{tab:unique_job_stats} presents the statistics of the normalized job titles. As shown in the table, the Singapore and Hong Kong datasets have 11.4\% and 10.8\% duplicate job titles respectively. On the other hand, the Switzerland dataset has very high duplicate job titles, 34.10\%. This may be attributed to the multilingual nature of the Switzerland data, containing many languages and different ways of naming job titles. After normalization, we finally have 29,084, 17,913 and 6,614 normalized job titles in Singapore, Switzerland and Hong Kong respectively.

\begin{table}[!t]
\caption{Statistics of normalized job titles (min\_sup $\geq$ 10)}
\label{tab:unique_job_stats}
\begin{tabular}{|l|r|r|r|}
\hline
Statistics & Singapore & Switzerland & Hong Kong \\
\hline
Number of parsed job titles 	& 32,828 & 27,184 & 7,415\\
Number of duplicate parsed job titles & 3,744 (11.4\%) & 9,271 (34.1\%)& 801 (10.8\%)\\
Number of normalized job titles & 29,084 & 17,913 & 6,614\\
\hline
\end{tabular}
\end{table}

\subsection{Network Construction}
\label{sub:network-construction}

We use the extracted job hops and the normalized job titles to form our talent flow network. A talent flow network is a directed graph where each edge represents a job hop activity. Based on node type in the network, talent flow network can be classified into two types: (1) \textbf{job network}, where each node $v_{t,i}$ represents a canonical job title $t$ in industry $i$, and (2) \textbf{organization network}, where each node $v_{c}$ represents an organization $c$. Job network allows us to observe talent flow at job level, whereas organization network allows us to observe talent flow at organization level.

For the job network, a directed edge from node $v_{t,i}$ to node $v_{t',i'}$ represents a job hop activity from a node $(t,i)$ to another node $(t',i')$. We also capture the number of user profiles moving from $(t,i)$ to $(t',i')$ as the \textbf{edge weight} $e_{(t,i) \rightarrow (t',i')}$. The same applies to the organization network, i.e., the edge weight $e_{c \rightarrow c'}$ represents the number of users moving from an organization $c$ to another organization $c'$.

%% file: method.tex
\section{Talent Flow Analytics}
\label{sec:analytics_methods}

This section elaborates the three types of talent flow analytics as shown in Fig.~\ref{fig:analytics-overview}, namely: (a) hop classification and analysis, (b) job attribute analysis, and (c) connectivity analysis.

\subsection{Hop Classification and Analysis}
\label{sub:hop-classification}

The hop classification and analysis essentially involve two types of job hop:
\begin{itemize}
\item \textbf{External hop}. This hop refers to transition from one job to another job, where the source and destination companies are \emph{different}. That is, an external hop is a hop from job $j = (t,c,i)$ to job $j' = (t',c',i')$, where $c \neq c'$. By this definition, the origin job title $t$ need not be the same as the destination title $t'$. Intuitively, two jobs with the same title but at different companies should be treated as separate jobs.

\item \textbf{Internal hop}. This is transition from one job to another, where the source and destination companies are \emph{the same}, i.e., an internal hop is a hop from job $j = (t,c,i)$ to job $j' = (t',c',i')$, where $c = c'$ and $t \neq t'$. The latter constraint ($t_{j} \neq t_{j'}$) is meant to avoid job title duplicates under the same company (e.g., a person may state three times that (s)he is a civil engineer at company X, for (s)he has worked on three construction projects under the same company). As such, we do \emph{not} count a move from job $j$ to $j'$ where $t = t'$ and $c = c'$ as a (valid) internal hop. 
\end{itemize}

\subsection{Job Attribute Analysis}
\label{sub:jobattribute-analysis}

In our study, we want to tell how much career advancement people make in their jobs.  We therefore first need to estimate the experience of a person holding a job. Secondly, to determine changes of job market over time, we need to estimate how long a job has existed.
To fulfill the two goals, we introduce several key metrics that are applied to the core users, i.e., those with at least one entry in the education and skill fields. With the two fields, one can derive interesting attributes of jobs and the skill profiles of the three economies. We describe the key metrics in turn below.

\begin{itemize}
  \item \textbf{Work experience}: This refers to the duration since the graduation date of the most recent educational degree of a person  till the time at which (s)he finishes a particular job. For a person $p$ with job title $t$ at organization $c$, the work experience is:
    \begin{align}
    wk\_exp(p, t, c, i) = end\_time(p, t, c, i) -
    grad\_date(p)
    \end{align}
where $grad\_date(p)$ denotes the last graduation date as mentioned in his/her account profile. In our subsequent analyses, note that we consider only positive work experience, i.e., we exclude cases whereby $wk\_exp(p, t, c, i) <= 0$. This is due to the observation that most jobs taken prior to the last education in our data are typically of interim nature (such as  internship), which may introduce bias in our analysis at higher (e.g., industry) level. Next, for a given job title $t$ in industry $i$, the average (i.e., expected) work experience of the job title-industry pair $(t,i)$ is given by:
    \begin{align}
    avg\_wk\_exp(t, i) = \frac{1}{|\mathbf{S}_{t,i}|} \sum_{(p,t,c,i) \in \mathbf{S}_{t,i}} wk\_exp(p,t,c,i)
    \end{align}
where $\mathbf{S}_{t,i}$ is the set of (unique) person-job pairs having job title $t$ in industry $i$. Examples of job title with high $avg\_wk\_exp$ score across industries in our data are ``Professor'', ``Managing Director'', and ``CEO'', whereas examples with low $avg\_wk\_exp$ score are ``Intern'' and ``Teaching Assistant''.

  \item \textbf{Job age}: This is the duration from the start of a given job until the current date $curr\_date$. It measures how recent or established a job is from the perspective of a person holding the job. For a person $p$ with job title $t$ at organization $c$, the job age is defined as:
    \begin{align}
    job\_age(p, t, c, i) = curr\_date - start\_date(p, t, c, i)
    \end{align}
    where $start\_date(p,t,c,i)$ refers to the start date of the person $p$'s job title $t$ at organization $c$ of industry $i$.
    For a given job title $t$ from industry $i$, the average (expected) age of the (job title, industry) pair $(t,i)$ is therefore:
    \begin{align}
    \label{eqn:job_age}
    avg\_job\_age(t, i) = \frac{1}{|\mathbf{S}_{t,i}|} \sum_{(p,t,c,i) \in \mathbf{S}_{t,i}} job\_age(p,t,c,i)
    \end{align}
    Examples of job titles with high $avg\_job\_age$ score across industries in our data are ``Director'', ``Systems Engineer'', and ``Division Manager'', while examples with low score are ``Data Scientist'' and ``Media Analyst''.

\end{itemize}

Based on the above metrics, we further derive several higher-level metrics, by aggregating over user profiles at either the job or organization level:
\begin{itemize}
\item \textbf{External hop fraction}: The fraction of people who move out from a organization $c$ to a different organization $c' \neq c$ over the (total) people hopping from organization $c$. Formally, for a given group of users $g$ (e.g., work experience, job age, or skill count group),job title translation and parsing the external hop fraction is:
\begin{align}
\label{eqn:ext_hop_frac}
\%external\_hop(g) = \frac{ | \mathbf{P}_{c \rightarrow c'}^g | }{ | \mathbf{P}_{c \rightarrow c'}^g | + | \mathbf{P}_{c \rightarrow c}^g | }
\end{align}
where $\mathbf{P}_{c \rightarrow c'}^g$ is the set of all user profiles belonging to group $g$ who perform \emph{external hops} from some arbitrary organizations $c$ to \emph{different} organizations $c' \neq c$. Conversely, $\mathbf{P}_{c \rightarrow c}^g$ is the set of user profiles belonging to group $g$ who perform \emph{internal hop} within the \emph{same} organization $c$.

\item \textbf{Job level}: As different organizations offer jobs of different rewards and seniority levels (even for the same job titles), we want to be able to measure them.  Since our data do not carry any salary information, we estimate the seniority level of a job $(t,c)$ by computing the \emph{average work experience} over all users who mention job title $t$ at organization $c$ in their profiles:
    \begin{align}
    \label{eqn:job_level}
    job\_level(t,c) &= \frac{1}{|\mathbf{P}_{t,c}|} \sum_{p \in \mathbf{P}_{t,c}} wk\_exp(p,t,c,i)	\end{align}
    where $\mathbf{P}_{t,c}$ is the set of all people who include job $(t,c)$ in their profiles. In the equation, $i$ can be inferred from $c$.  Intuitively, a job with longer average work experience implies that a longer time is required to achieve that position, and hence we can expect it to be a high-level job (e.g., CEO of a multi-national organization).

\item \textbf{Level gain}: This refers to the difference between the levels of two jobs within the same or different companies. A positive level gain can be loosely interpreted as a ``promotion'', whereas a negative level gain loosely implies a ``demotion''. Here the ``promotion'' (``demotion'') does not necessarily mean a monetary increase (decrease), but more of an increase (decrease) in the level of work experience required. Formally, the level gain for hop from job $(t,c)$ to job $(t',c')$ is given by:
    \begin{align}
    level\_gain((t,c), (t',c')) = job\_level(t',c') - job\_level(t,c)
    \end{align}
    We note that, although there is no ground truth available in our OPN data, our manual inspections show that the level gain provides a reasonable proxy for a promotion or demotion. It is also worth mentioning that no zero level gain (i.e., neither ``promotion'' or ``demotion'') is found in our data.
\end{itemize}

\subsection{Connectivity Analysis}
\label{sub:network-analysis}

To facilitate connectivity analysis, we utilize several network centrality metrics to measure node importance in both job and organization networks, as follows:
\begin{itemize}
\item \textbf{In-degree centrality}. This metric refers to the number of inbound (unweighted) edges for a node in the job or organization graph. The in-degree centrality can be interpreted as a measure of how \emph{prominent} a job (or organization) is in a local sense---a high in-degree may imply that it attract talents from the immediate in-neighbors. For this metric, we do not take into account the edge weight information (i.e., the total number of incoming user profiles), as we want to minimize the support bias due to a large number of users for a given job (organization).

\item \textbf{Out-degree centrality}. This is defined as the number of outbound (unweighted) edges for a node in the job or organization graph. We can use the out-degree centrality to measure how \emph{influential} a job (or organization) is in a local sense---a high out-degree may be indicative of a talent supplier to the immediate out-neighbors. Again, we do not utilize the edge weight to compute this metric, so as to mitigate the support bias.

\item \textbf{PageRank centrality}. This is a well-known metric originally used to rank web pages \cite{Brin1998}. PageRank views inbound edges as ``votes'', and the key idea is that ``votes'' from important nodes should carry more weight than ``votes'' from less important nodes. In this work, we employ a \emph{weighted} version of PageRank \cite{Langville2005}, whereby the transition probabilities for each (source) node is proportional to the (out-)edge weights divided by the weighted out-degree of the node. In the context of job and organization graphs, the weighted PageRank can be viewed as a measure of \emph{global competitiveness}---a job or organization with high PageRank reflects a ``desirable'' destination point where the flow of talent is heading to. Here we use edge weight, as the hop volume matters in determining where the flow goes to. To avoid dead ends (i.e., nodes with zero out-degree), we allow our PageRank to perform random jump with the default ``teleportation'' probability of $0.15$.
\end{itemize}

%% file: insights.tex
\section{Insights and Discussion}
\label{sec:insights}

Using the methodology and metrics described previously in Sections \ref{sec:overview}--\ref{sec:analytics_methods}, we present our empirical findings and analysis in this section. We begin our discussion with basic distribution analysis, followed by findings in each type of analysis within the talent flow analytics phase.

\subsection{Distribution Analysis}

\begin{figure}[!t]
\captionsetup[subfigure]{labelformat=empty}
\centering
\subfloat[]{
  \includegraphics[width=0.32\columnwidth]{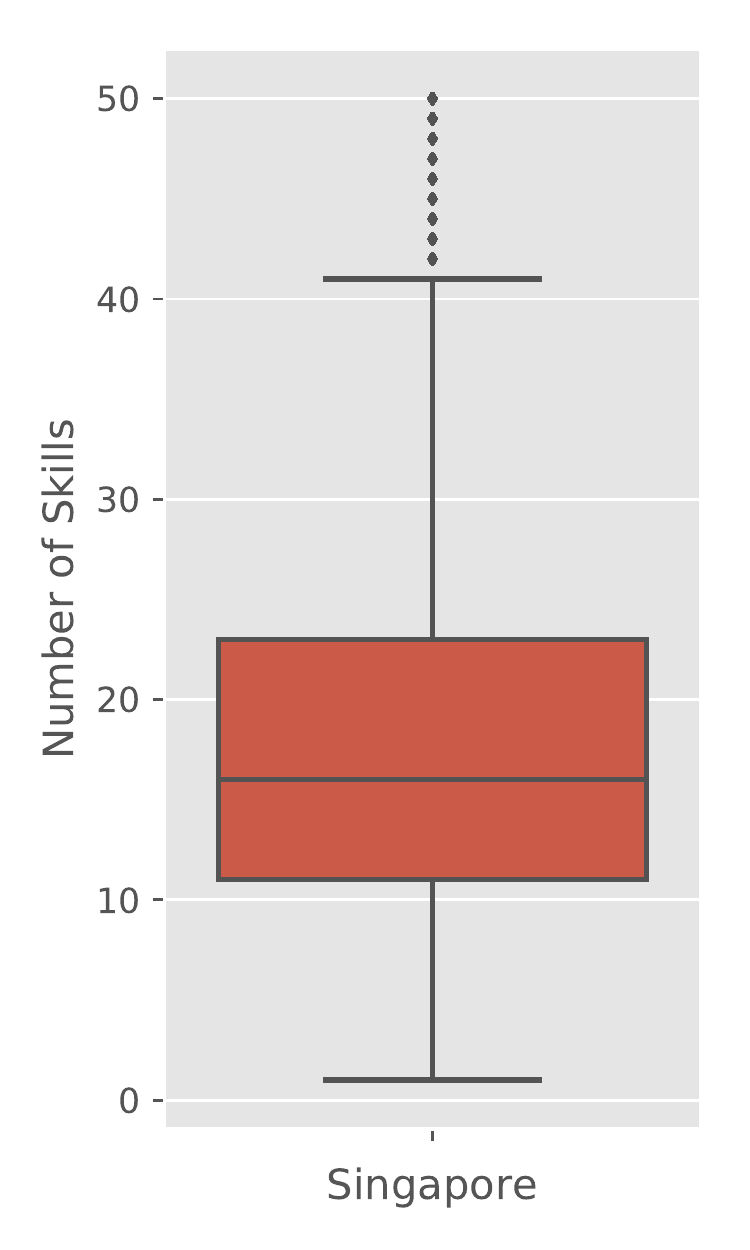} 
}
\subfloat[]{
  \includegraphics[width=0.32\columnwidth]{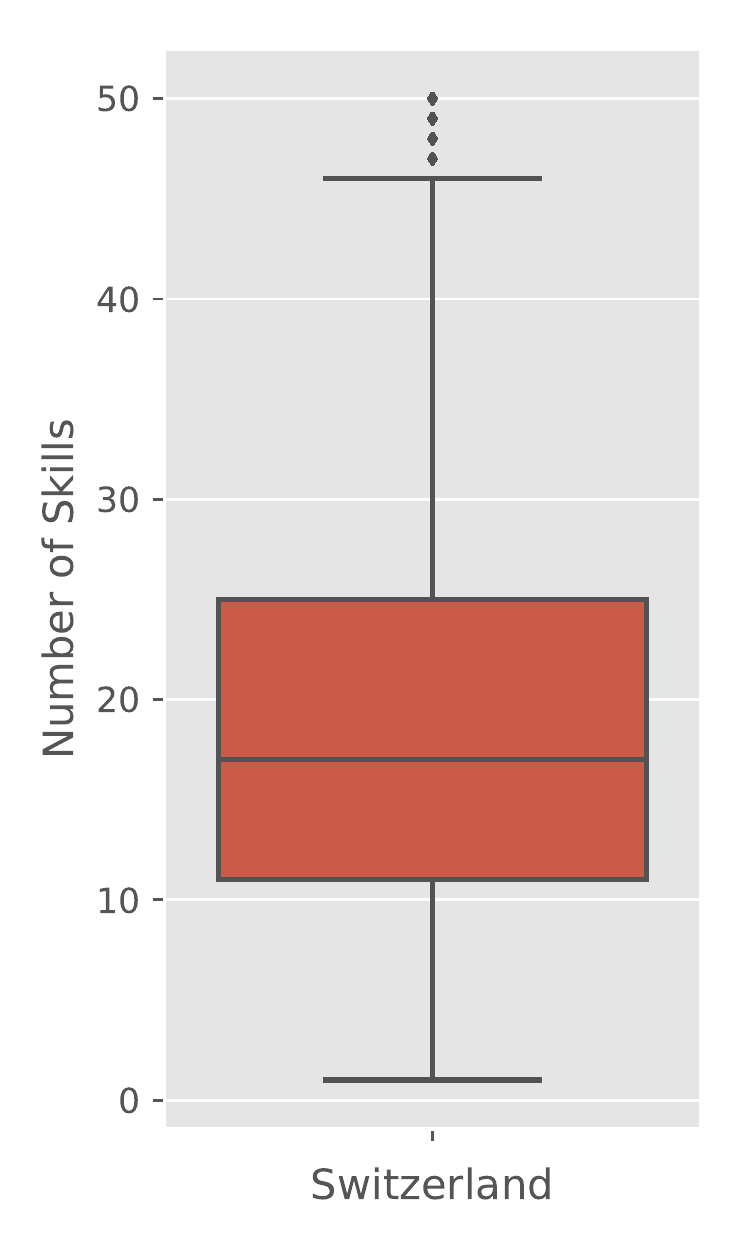} 
}
\subfloat[]{
  \includegraphics[width=0.32\columnwidth]{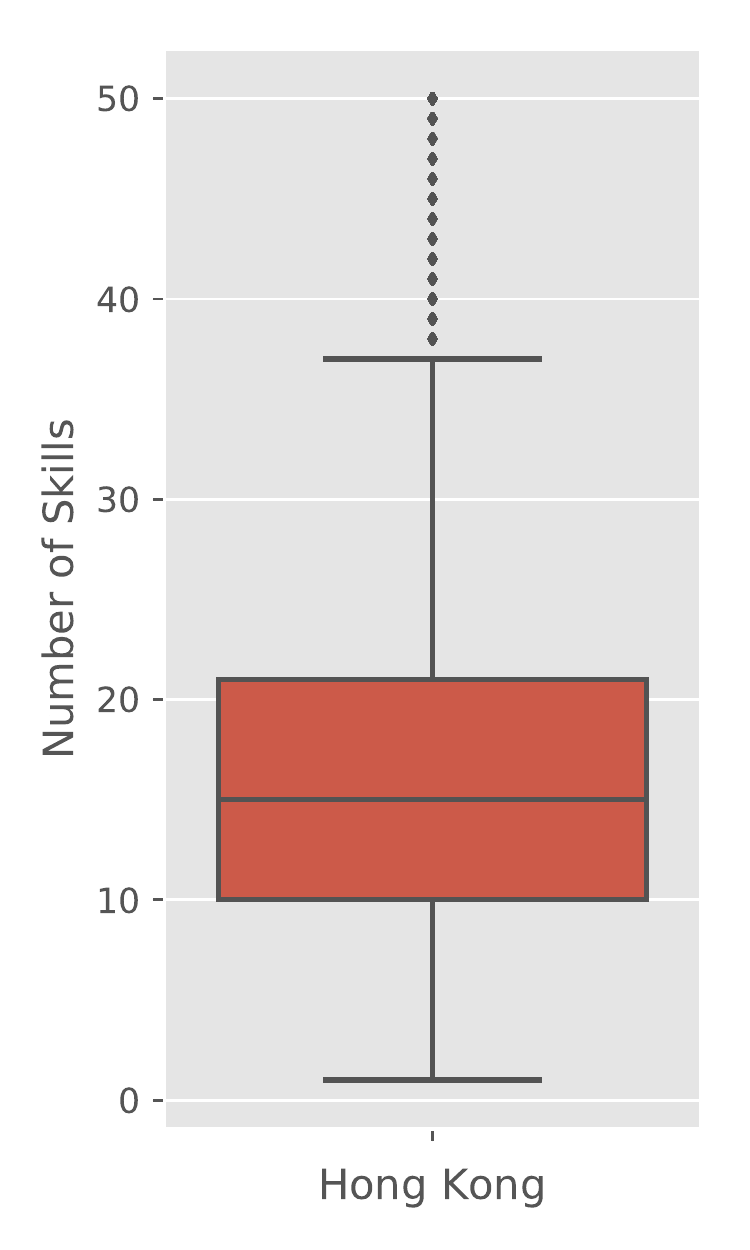} 
}
\caption{Distribution of number of skills}
\label{fig:num_skills_boxplot}
\end{figure}

We first analyze the distributions of several basic metrics, including skill count, work experience, and job age. We found that the \emph{core user profiles} in our talent flow networks typically have about 10-25 skills. Fig.~\ref{fig:num_skills_boxplot} shows the box plots of the skill distribution. It is shown that Switzerland workforce tends to have (or rather declare) slightly more skills compared to Singapore and Hong Kong workforce. It also indicates that Hong Kong workforce has (or declares) the least number of skills. The maximum of 50 skills is due to the fact that our OPN imposes a maximum limit of 50 skills per user profile. 

We also notice that most jobs in Singapore, Switzerland, and Hong Kong consist of young workforce, which has work experience of 5 years or less. This pattern holds for all the three datasets. Most users in our OPN data are relatively young in terms of work experience. This could be due to the younger users showing more interest in using OPN to conduct professional networking. On the other hand, there are only very few people who have worked for over 20 years. The most common work experience (i.e., the mode) is 2 years for Singapore and Hong Kong, and 1 year for Switzerland. 

\begin{figure}[!t]
\centering
\subfloat[Singapore]{
  \includegraphics[width=0.66\columnwidth]{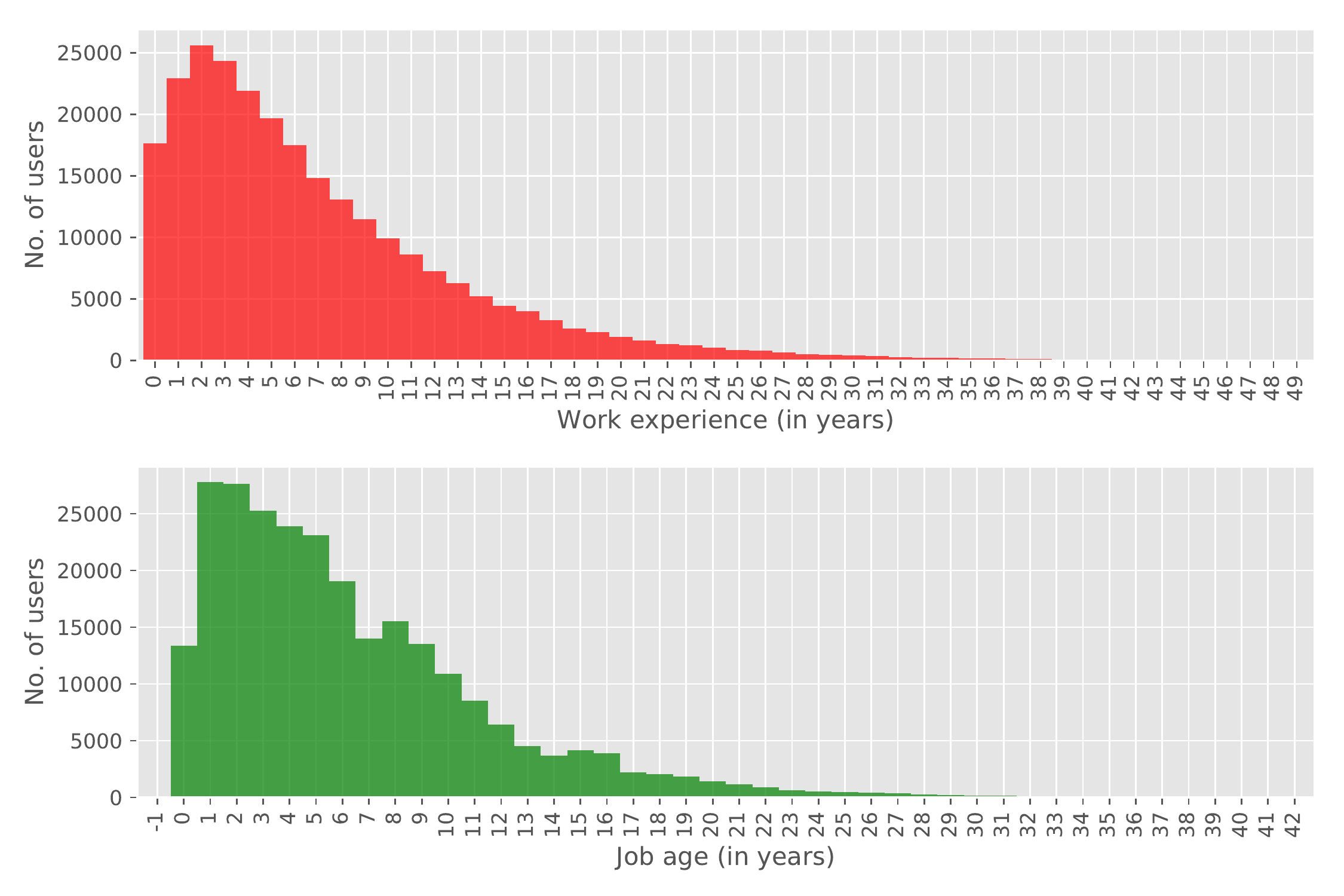} 
}\vfill
\subfloat[Switzerland]{
  \includegraphics[width=0.66\columnwidth]{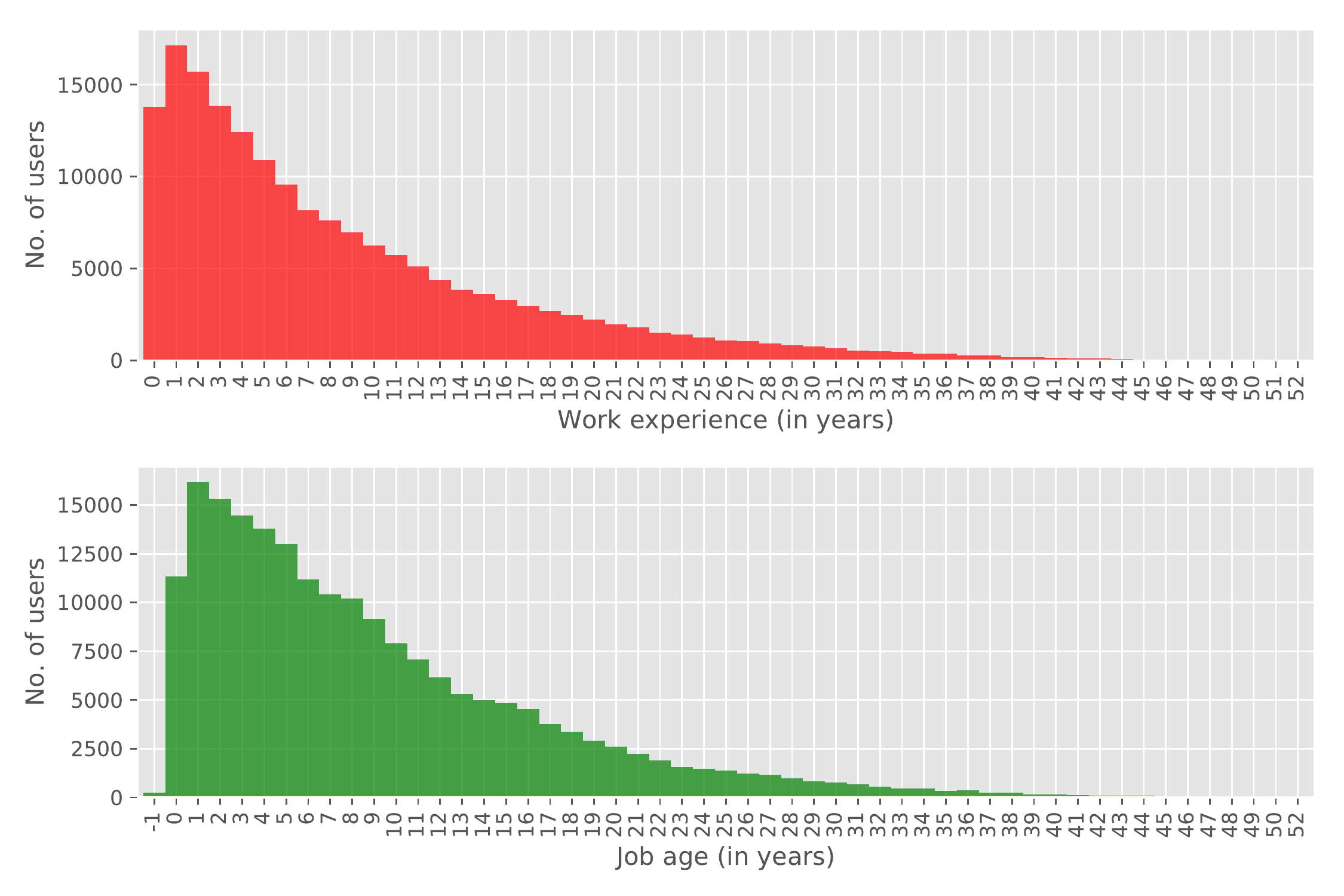} 
}\vfill
\subfloat[Hong Kong]{
  \includegraphics[width=0.66\columnwidth]{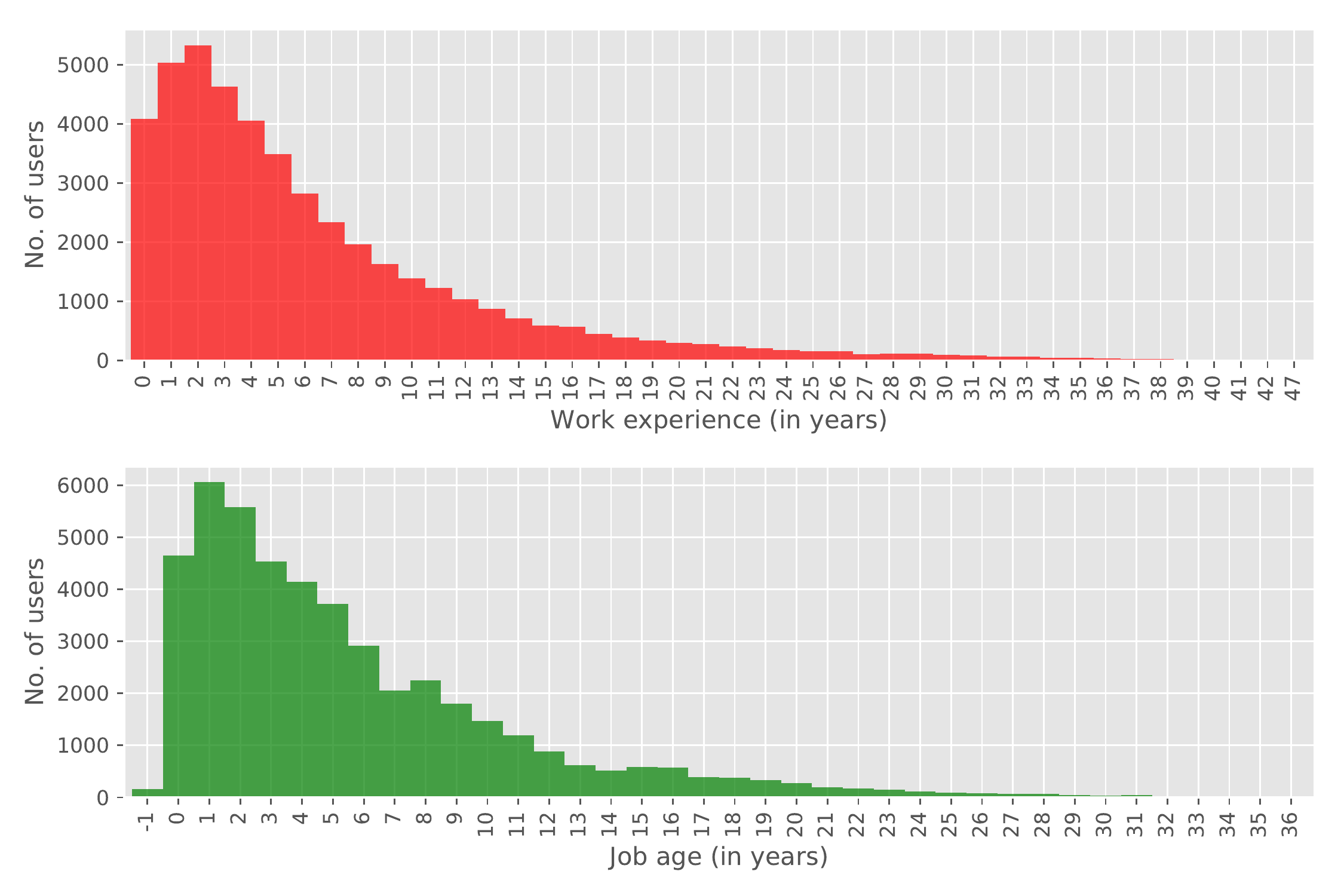} 
}
\caption{Distributions of work experience and job age}
\label{fig:job_dist}
\end{figure}

Fig.~\ref{fig:job_dist} presents the distributions of the work experience and job age across the three countries. The results suggest that most jobs have been established for 1 year or more. On the other hand, only very few jobs have been established for more than 20 years. As with work experience, the most common job age is 1 year. The relatively young job age can be explained partly by the young user base, and partly by the sparsity of old but senior-level jobs. From the labor economics perspective, this suggests that attention need to be given to identifying and creating more senior jobs to support an aging workforce.

\begin{figure}[!t]
\centering
\subfloat[Singapore]{
  \includegraphics[width=0.66\columnwidth]{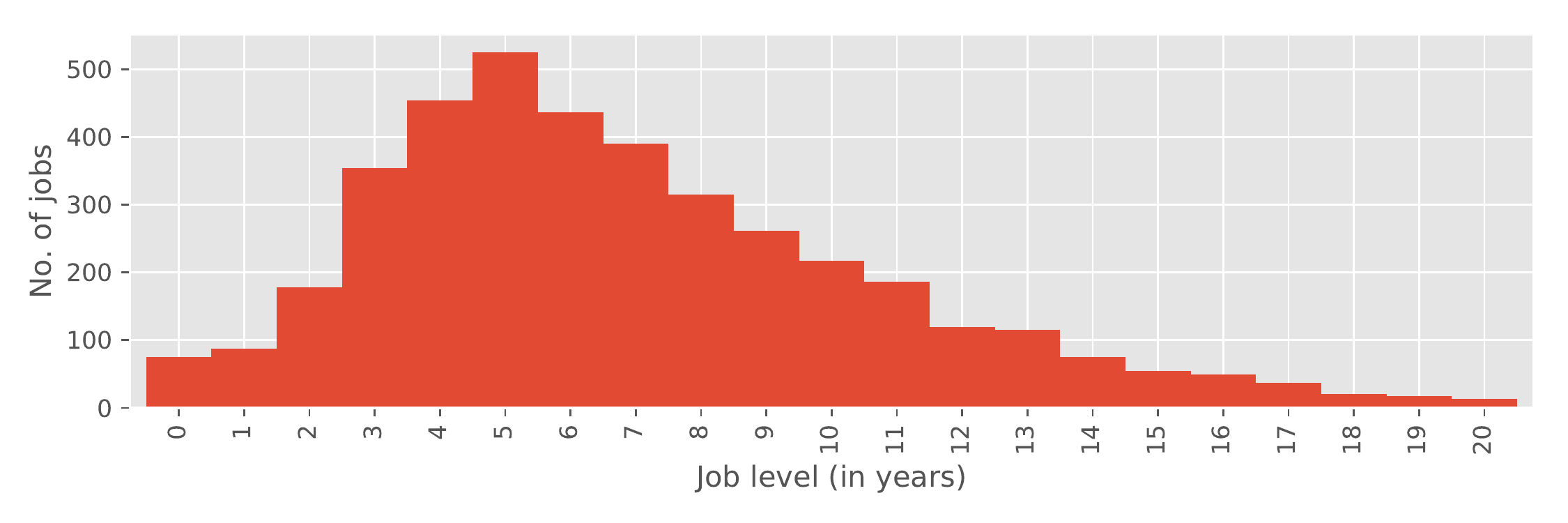} 
}\vfill
\subfloat[Switzerland]{
  \includegraphics[width=0.66\columnwidth]{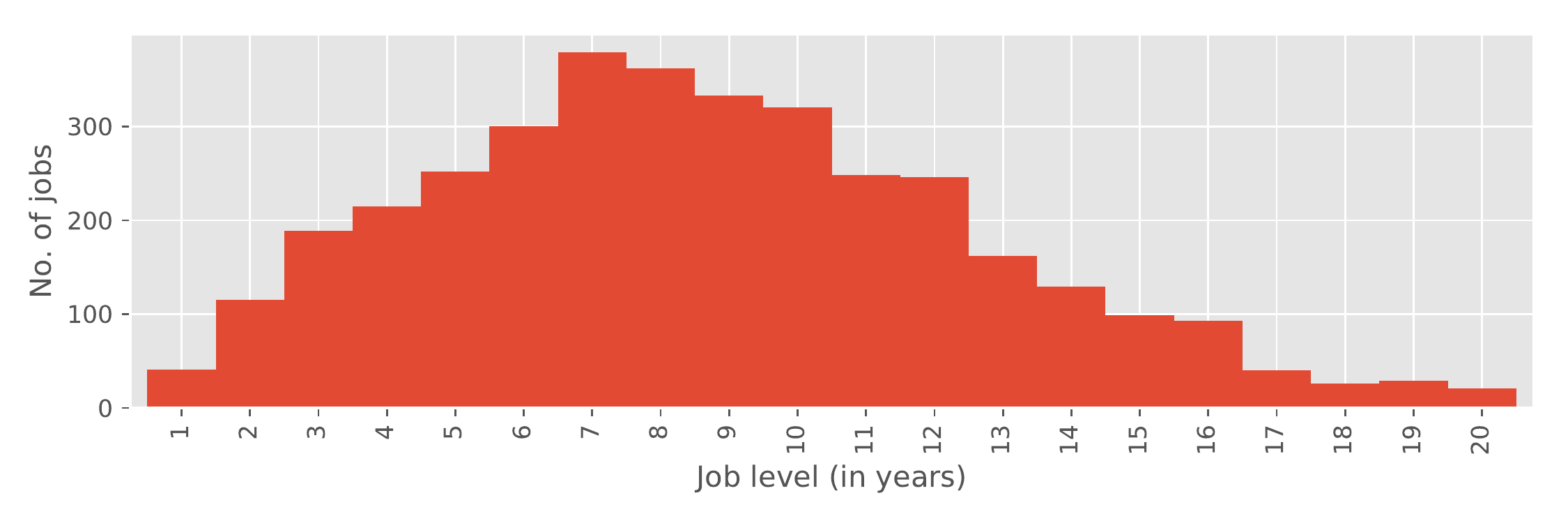} 
}\vfill
\subfloat[Hong Kong]{
  \includegraphics[width=0.66\columnwidth]{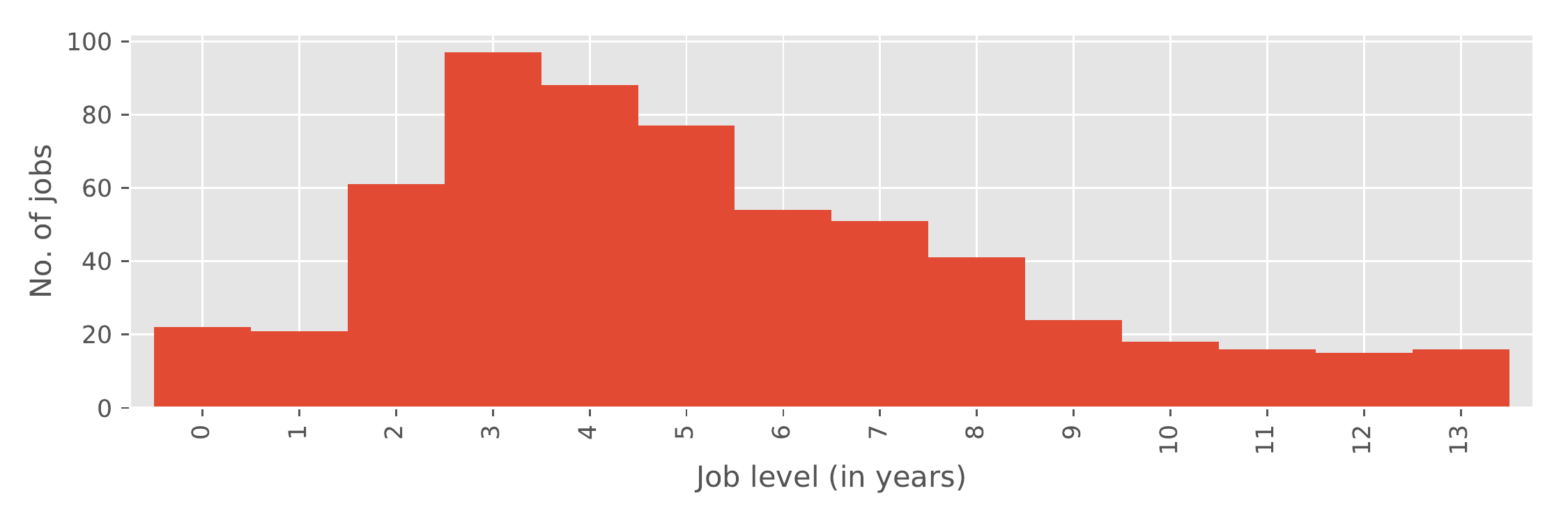} 
}
\caption{Distribution of job level}
\label{fig:job_level_dist}
\end{figure}

The job level distribution shown in Fig.~\ref{fig:job_level_dist}(a) reveals that most jobs in Singapore possess 4-6 years of experiences. Compared to jobs in Singapore, most jobs in Switzerland seem to have more senior jobs having 7-8 years of experiences (cf. Fig.~\ref{fig:job_level_dist}(b)), and most jobs in Hong Kong seem to have more junior jobs with 3-4 years of experiences (cf. Fig.~\ref{fig:job_level_dist}(c)). We also notice that the distribution of job level in Hong Kong is much shorter than that of Singapore and Switzerland and the maximum job level is 13 years. This is understandable nevertheless, considering the fact that Hong Kong began as a new special administration of China only after 1997.

\begin{figure}[!t]
\centering
\subfloat[Singapore vs. Switzerland]{
  \includegraphics[width=0.66\columnwidth]{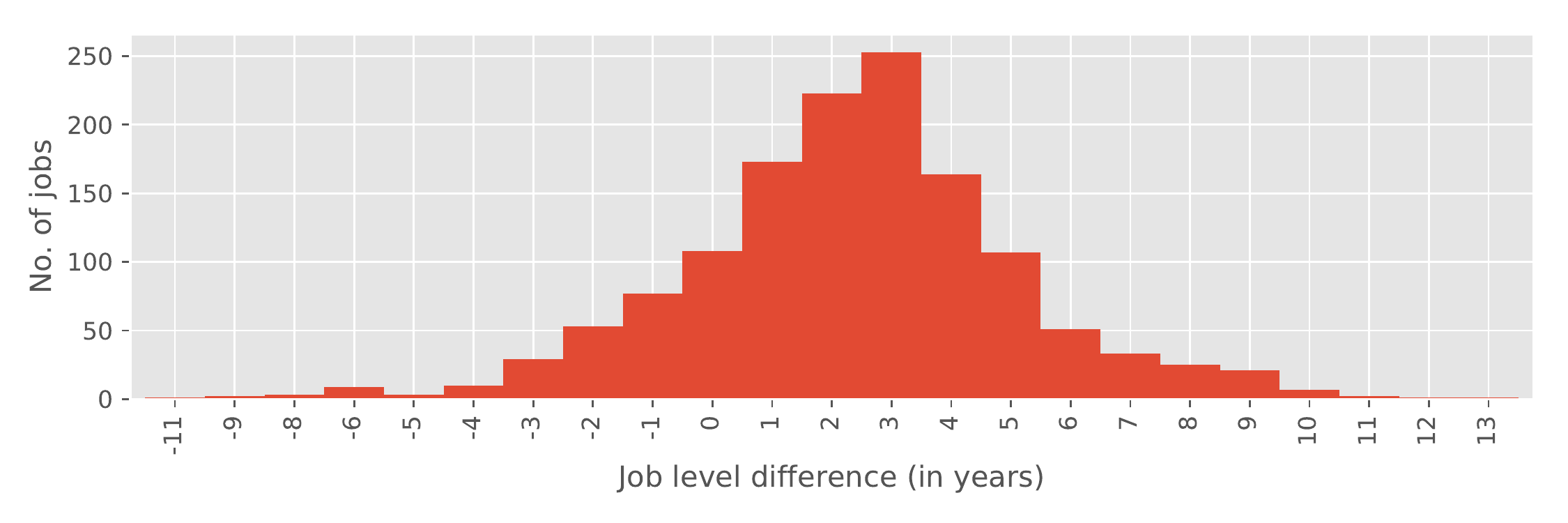} 
}\vfill
\subfloat[Singapore vs. Hong Kong]{
  \includegraphics[width=0.66\columnwidth]{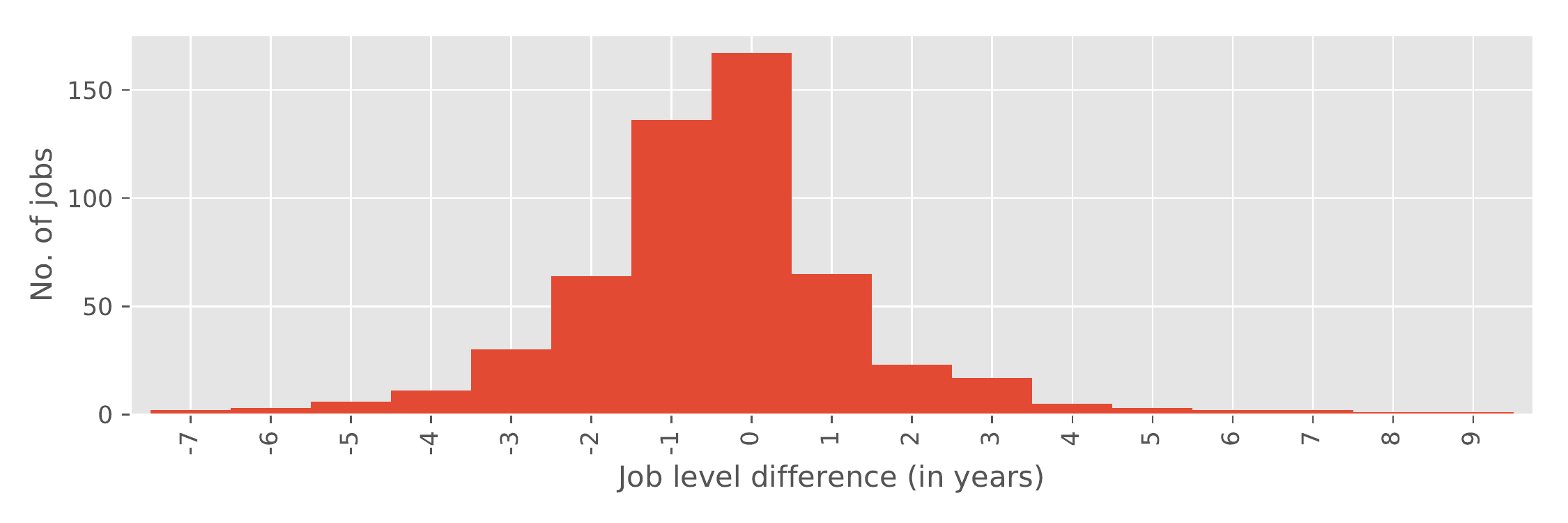} 
}
\caption{Discrepancy of job level}
\label{fig:job_level_diff}
\end{figure}

\begin{figure}[!t]
\centering
\subfloat[Singapore vs. Switzerland]{
  \includegraphics[width=0.66\columnwidth]{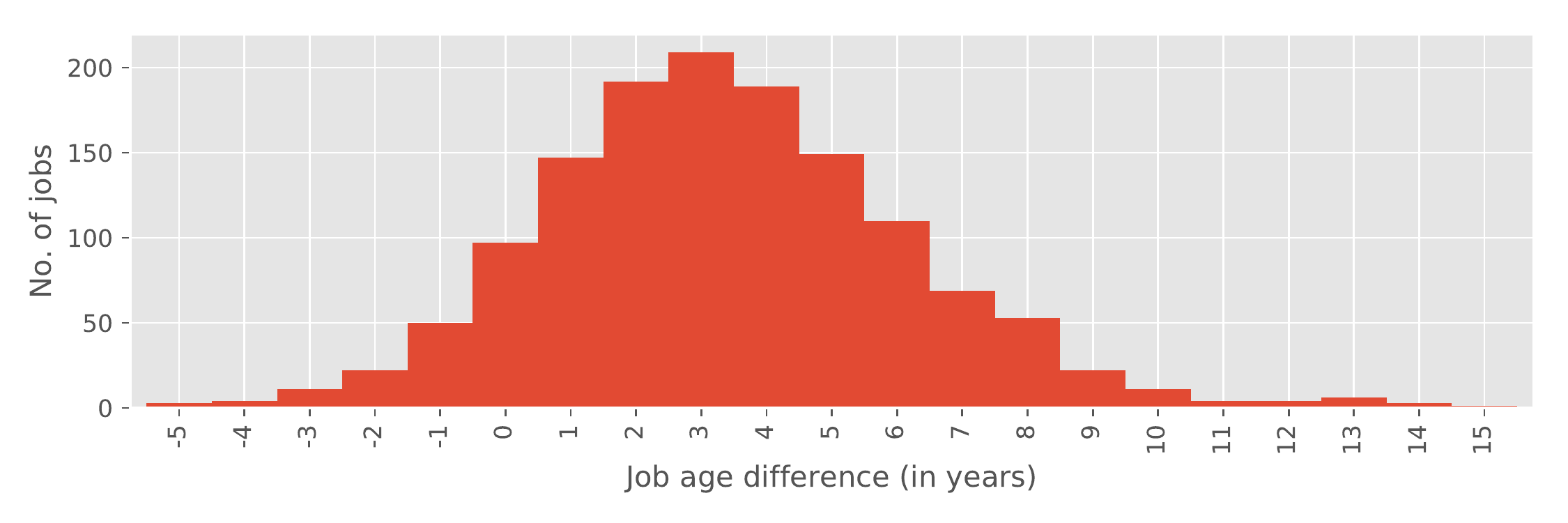} 
}\vfill
\subfloat[Singapore vs. Hong Kong]{
  \includegraphics[width=0.66\columnwidth]{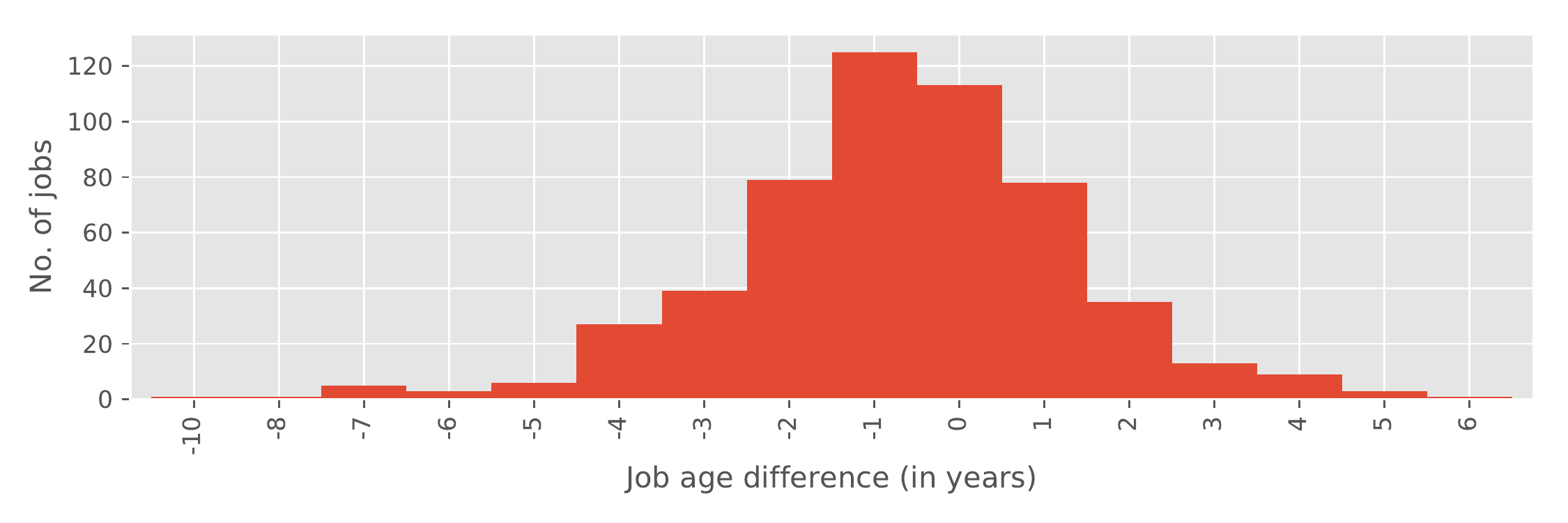} 
}
\caption{Discrepancy of job age}
\label{fig:job_age_diff}
\end{figure}

We conduct further investigation by examining the job level discrepancy of same job titles in Singapore vs. Switzerland as well as Singapore vs. Hong Kong. Fig.~\ref{fig:job_level_diff} summarizes the results, providing an interesting insight that explains the difference of job level distribution in Fig.~\ref{fig:job_level_dist}. We found that for the same job titles, jobs in Switzerland tend to have 2-3 longer years of experiences than jobs in Singapore, whereas jobs in Hong Kong tend to have equally long (or slightly shorter) years of experiences than jobs in Singapore.

Comparing Fig. \ref{fig:job_dist} and \ref{fig:job_level_dist}, it can be observed that the distributions of job age and job level are not identical. This is expected, based on the definitions of the two metrics in equations (\ref{eqn:job_age}) and (\ref{eqn:job_level}) respectively. In particular, job level looks \emph{forward} in time at how long an individual accumulates experience since his/her (last) graduation date, while job age looks \emph{backward} in time at how long a job has been established until a current reference time point.

Last but not least, comparison of the job age discrepancy for same job titles in Singapore vs Switzerland and Singapore vs Hong Kong tells us that most jobs in Switzerland tend to have longer job ages than those in Singapore. This implies that the Switzerland workforce is ahead of the Singapore counterparts in terms of job establishment for the same job title. On the other hand, Hong Kong workforce tends to have lower job age than Singapore workforce, implying that the jobs in Hong Kong are less established than in Singapore. Fig.~\ref{fig:job_age_diff} presents a summary of the job age comparison.

\subsection{Hop Classification and Analysis}
\label{sec:hop_analysis}

For this analysis, we started with an initial hypothesis that the propensity of external hop is potentially associated (correlated) with the work experience, job age, and number of skills. To test this, we conduct an investigation on how the external hop fraction (cf. equation (\ref{eqn:ext_hop_frac})) varies with different combinations of work experience, job age, and skill count groups. 

Fig.~\ref{fig:hop_fraction_grouped} shows the distribution of external hop fraction varying work experience, job age, and skill count group, whereby the minimum support was set to 100 for each bar in the plots. The figure reveals a number of key insights:

\begin{figure}[!t]
\centering
\subfloat[Singapore\label{fig:sg_hop_fraction_grouped}]{
  \includegraphics[width=1\columnwidth]{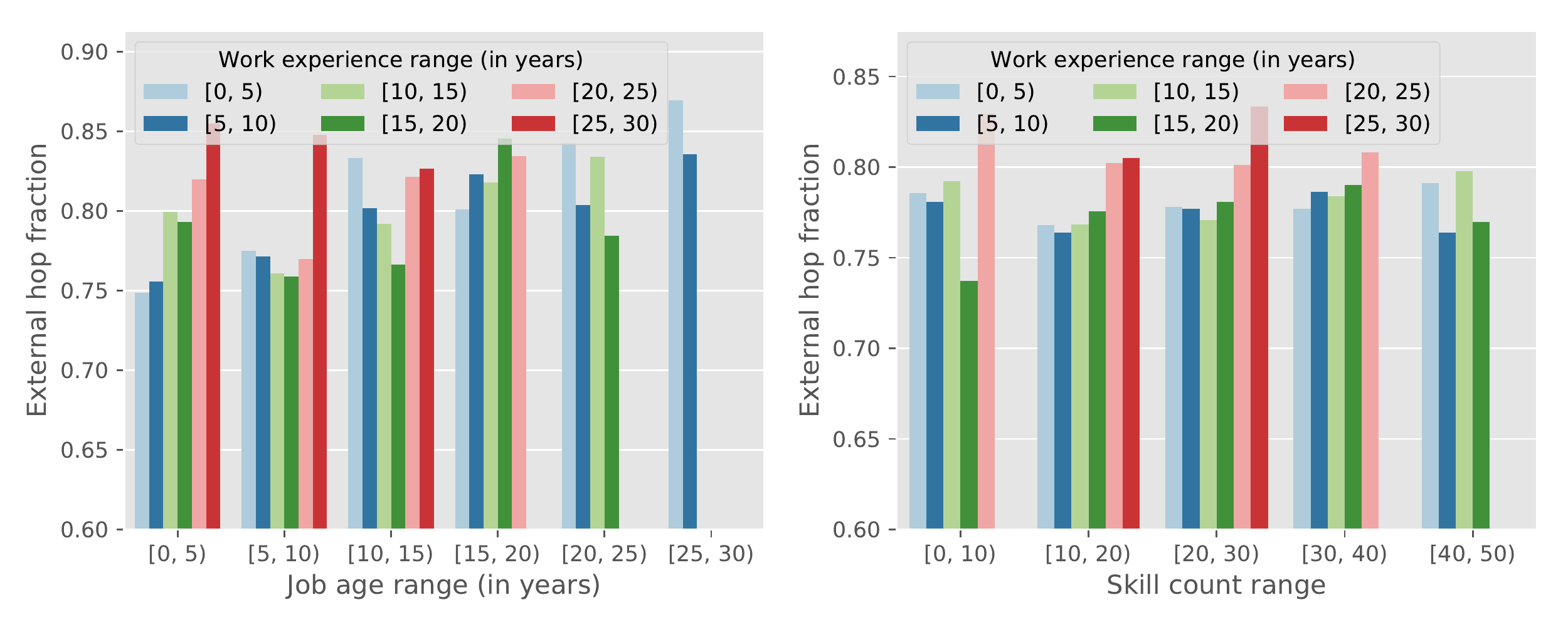} 
}\vfill
\subfloat[Switzerland\label{fig:ch_hop_fraction_grouped}]{
  \includegraphics[width=1\columnwidth]{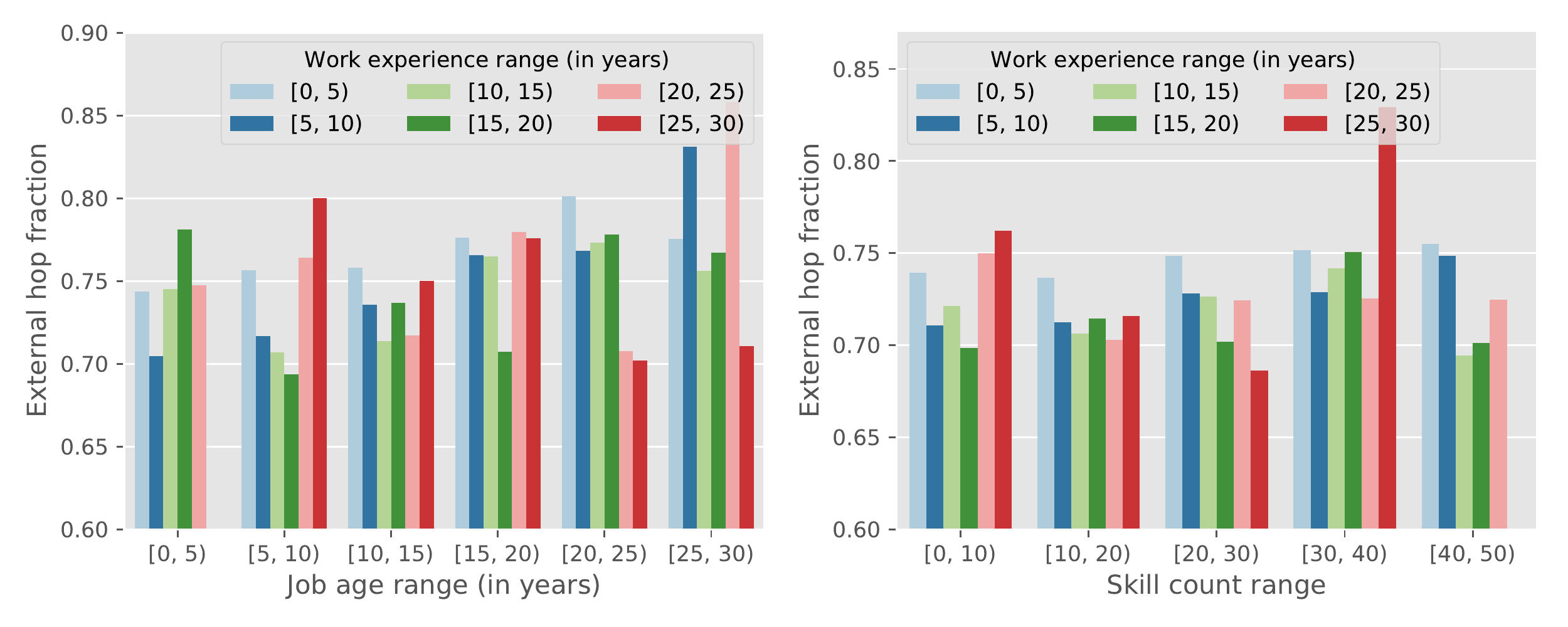} 
}\vfill
\subfloat[Hong Kong\label{fig:hk_hop_fraction_grouped}]{
  \includegraphics[width=1\columnwidth]{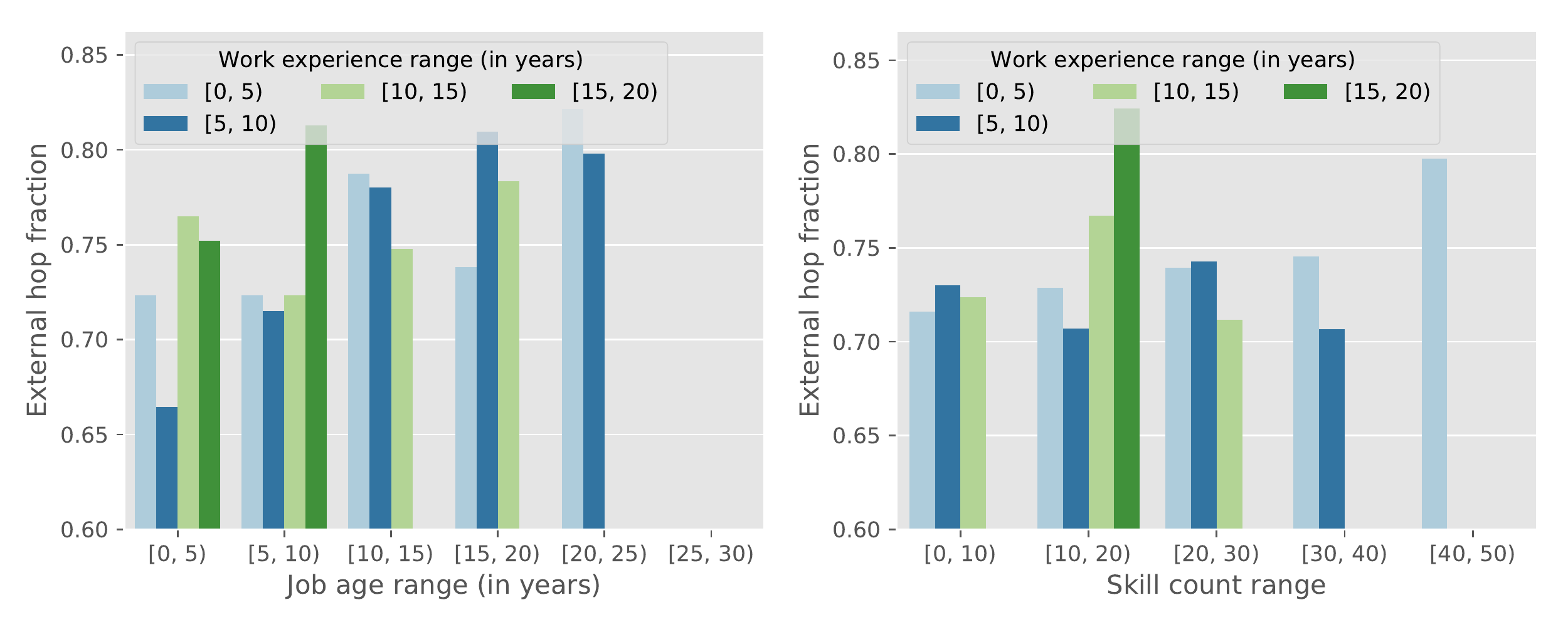} 
}
\caption{Distributions of external hop fraction}
\label{fig:hop_fraction_grouped}
\end{figure}

\begin{itemize}
\item External hops are generally very common in Singapore, Switzerland, and Hong Kong, regardless of work experiences, job age, and number of skills. The external hop fraction in Singapore, Switzerland, and Hong Kong are generally greater than 75\%, 70\%, and 65\% respectively. 
\item We found no strong evidence that external hops are influenced by work experience, job age, and number of skills, as no apparent patterns emerged on the charts. As such, we cannot accept our initial hypothesis. A plausible explanation is that other incentives that are unobservable from our data---such as monetary, work packages, and perks---might play more important roles in incentivizing external hops. Further investigations can be done by augmenting auxiliary information from other data sources from the relevant authorities, which is beyond the scope of our current work.
\item For the Hong Kong data, we do not observe any users with work experience $\geq$ 20 years. This conforms with our earlier finding in Fig.~\ref{fig:job_level_dist}, whereby the maximum job level found in the Hong Kong data is about 13 years. We can attribute this to the fact that Hong Kong is a relatively young special territory of China, which started in 1997.
\end{itemize}

Naturally, one can extend the above analysis at more granular levels, such as company and industry levels within a country. It is worth highlighting, however, that the data at these levels are typically sparse and/or noisy. Indeed, we have observed in our OPN data that many small companies and industries have low support (i.e., low number of people and/or jobs). This would lead to unreliable metrics/statistics computation that would prevent us from deriving meaningful insights. Handling data sparsity using a more sophisticated analytics approach is beyond the scope of this paper, but is certainly an avenue worthy further investigation in the future. 

\subsection{Job Attribute Analysis}

\begin{table}[!t]
\caption{Hop classification statistics for Singapore dataset}
\label{tab:sg_hop_type_stats}
\begin{tabular}{|l|c|c|c|c|}
\hline
     & Promotion & Demotion & Total \\
\hline
External hop & 2,627 & 930 & 3,557 \\
Internal hop & 2,182 & 200 & 2,382 \\
\hline
Total        & 4,809 & 1,130 & 5,939 \\
\hline
\end{tabular}
\end{table}

\begin{table}[!t]
\caption{Hop classification statistics for Switzerland dataset}
\label{tab:ch_hop_type_stats}
\begin{tabular}{|l|c|c|c|c|}
\hline
     & Promotion & Demotion & Total \\
\hline
External hop & 859 & 533 & 1,392 \\
Internal hop & 1,486 & 341 & 1,827 \\
\hline
Total        & 2,345 & 874 & 3,219 \\
\hline
\end{tabular}
\end{table}

\begin{table}[!t]
\caption{Hop classification statistics for Hong Kong dataset}
\label{tab:hk_hop_type_stats}
\begin{tabular}{|l|c|c|c|c|}
\hline
     & Promotion & Demotion & Total \\
\hline
External hop & 175 & 56 & 231 \\
Internal hop & 350 & 36 & 386 \\
\hline
Total        & 525 & 92 & 617 \\
\hline
\end{tabular}
\end{table}

\begin{figure}[!t]
\centering
\subfloat[Singapore]{
  \includegraphics[width=1\linewidth]{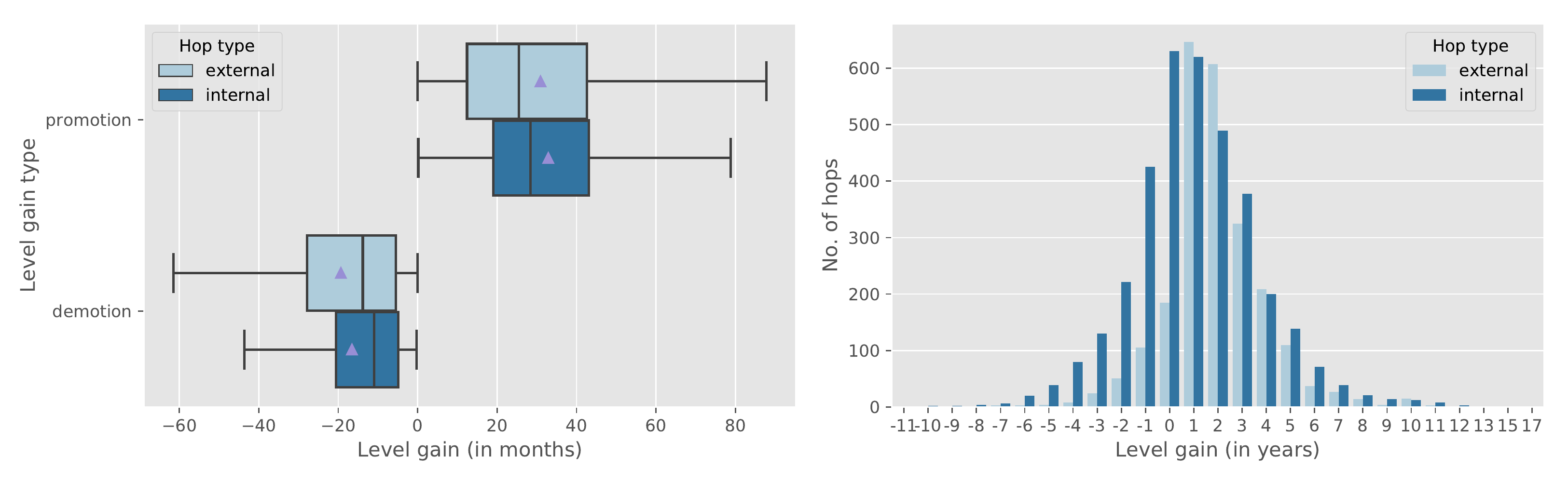} 
}\vfill
\subfloat[Switzerland]{
  \includegraphics[width=1\linewidth]{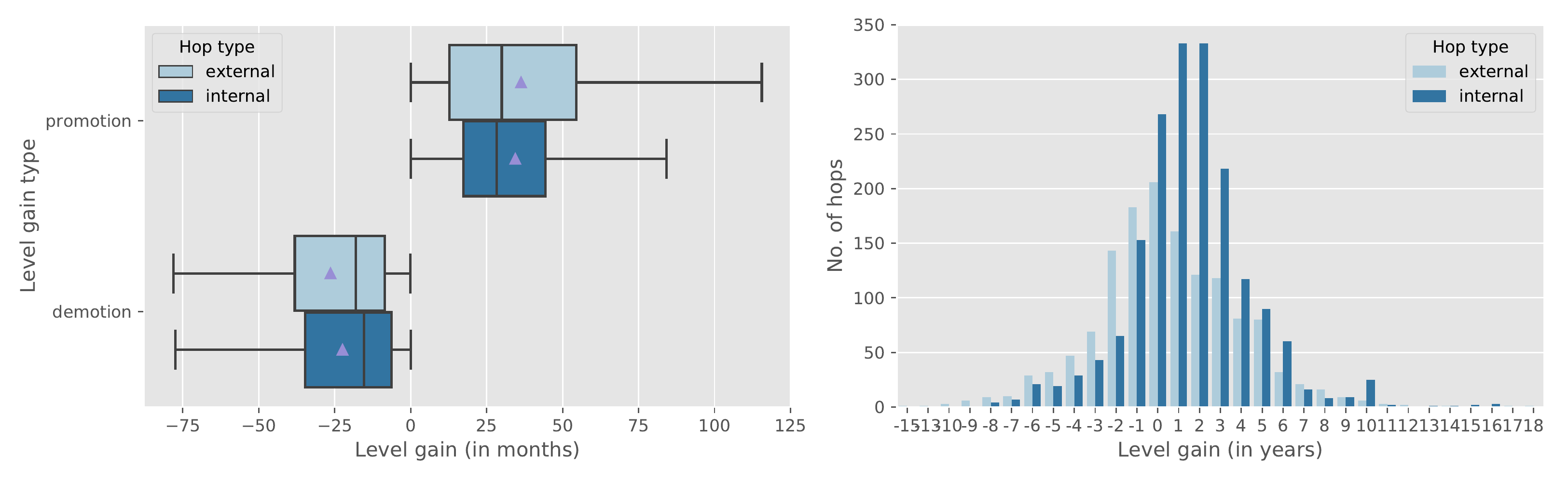} 
}\vfill
\subfloat[Hong Kong]{
  \includegraphics[width=1\linewidth]{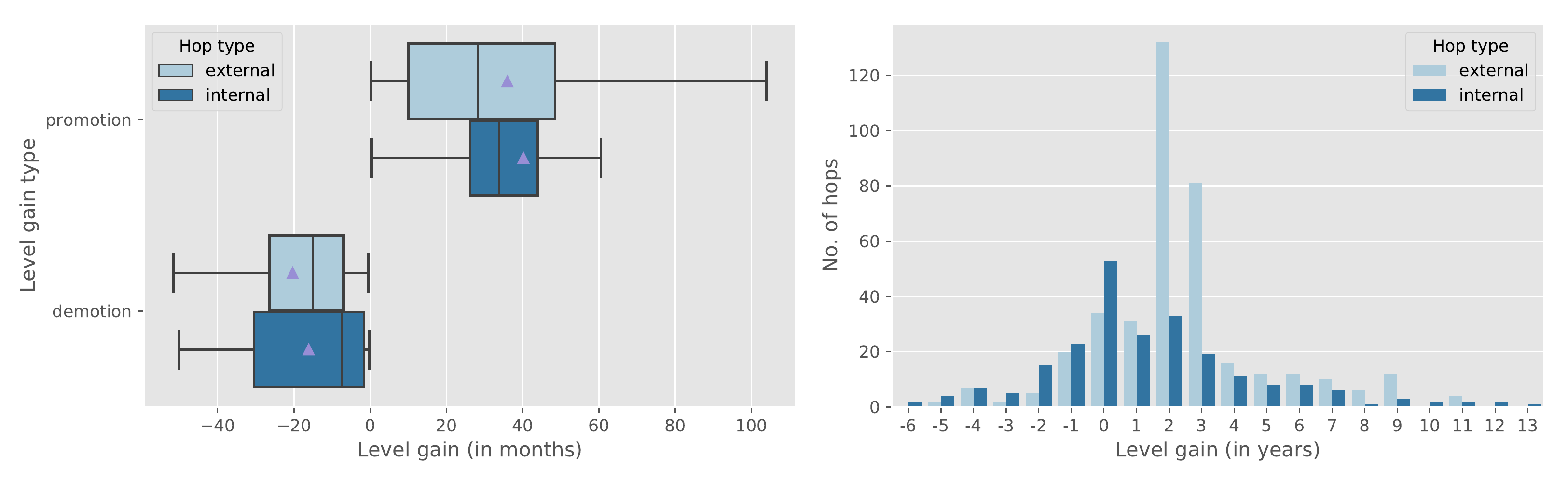} 
}
\caption{Comparison of level gains for different hop types}
\label{fig:hop_level_diff}
\end{figure}

As promotion is often a cited reason for people leaving one job for another, we now conduct a promotion and demotion analysis by dividing the hops into external and internal hops based on level gain (i.e., promotion vs. demotion). Table~\ref{tab:sg_hop_type_stats},~\ref{tab:ch_hop_type_stats}, and~\ref{tab:hk_hop_type_stats} show the statistics of promotion and demotion in Singapore, Switzerland, and Hong Kong respectively, and Fig.~\ref{fig:hop_level_diff} provides more detailed results. To get a reliable estimate of level gain---and in turn reliable “promotion” or “demotion” labels, we require both source and target jobs for each hop must fulfill the (default) minimum support of 10. As such, we do not include in the tables and figures hops that fail to meet the minimum support criterion. 

From the results in Tables~\ref{tab:sg_hop_type_stats},~\ref{tab:ch_hop_type_stats}, and~\ref{tab:hk_hop_type_stats}, we can derive several conclusions:
\begin{itemize}
\item The Singapore workforce does substantially more external hops than internal hops. In contrast, Switzerland and Hong Kong workforces perform less external hops than internal hops. This suggests that the Switzerland and Hong Kong workforces are generally more ``loyal'' than that of Singapore.
\item The probability of promotion is generally greater than that of demotion (for both external and internal hops across the three countries). Specifically, the Singapore dataset shows 81\% promotion as compared to 19\% demotion, Switzerland dataset shows 73\% promotion as compared to 27\% demotion, and Hong Kong dataset shows 85\% promotion and 15\% demotion. This matches the common intuition that a hop is more likely motivated by a job promotion than demotion.
\item The Singapore workforce generally tends to seek promotion via external hops, whereas Switzerland and Hong Kong people prefer to seek promotion via internal hops. This relates back to our first point, about the Switzerland and Hong Kong workforces being more loyal than that of Singapore.
\end{itemize}

\begin{figure}[!t]
\centering
\subfloat[Singapore]{
  \includegraphics[width=0.9\columnwidth]{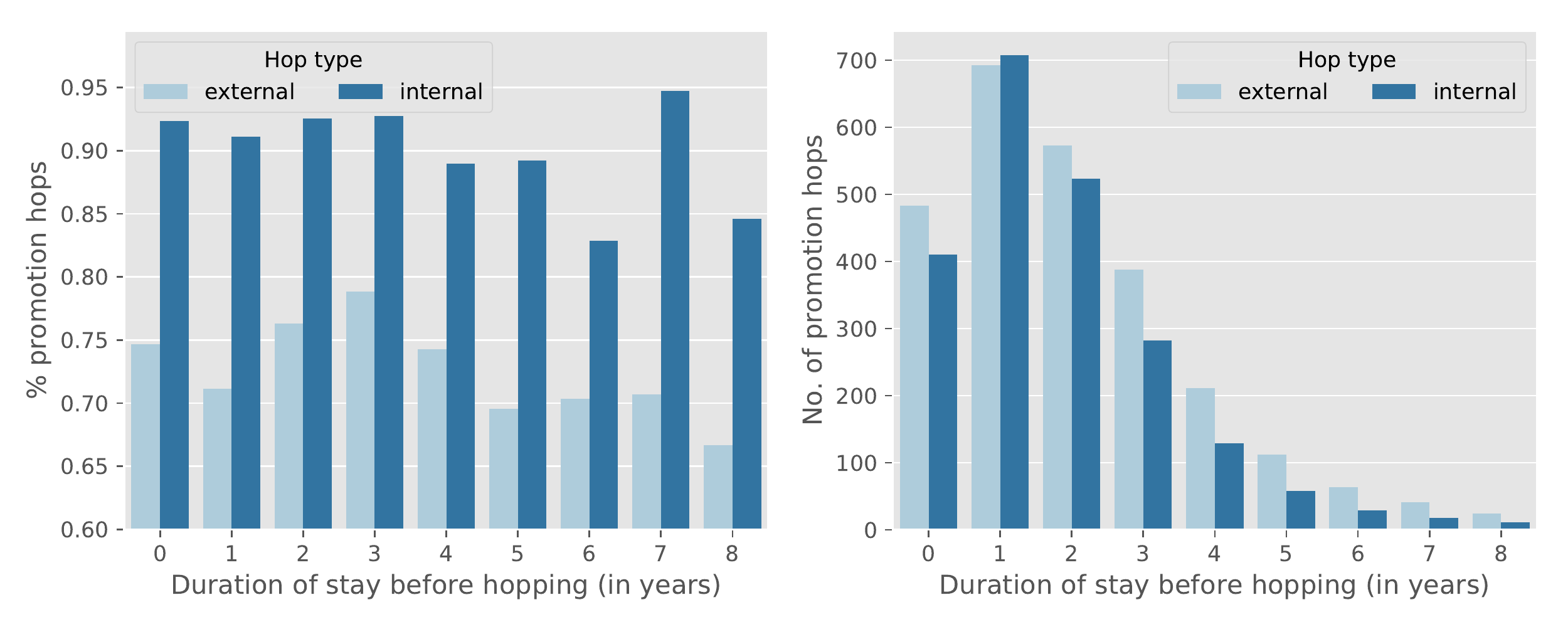} 
}\vfill
\subfloat[Switzerland]{
  \includegraphics[width=0.9\columnwidth]{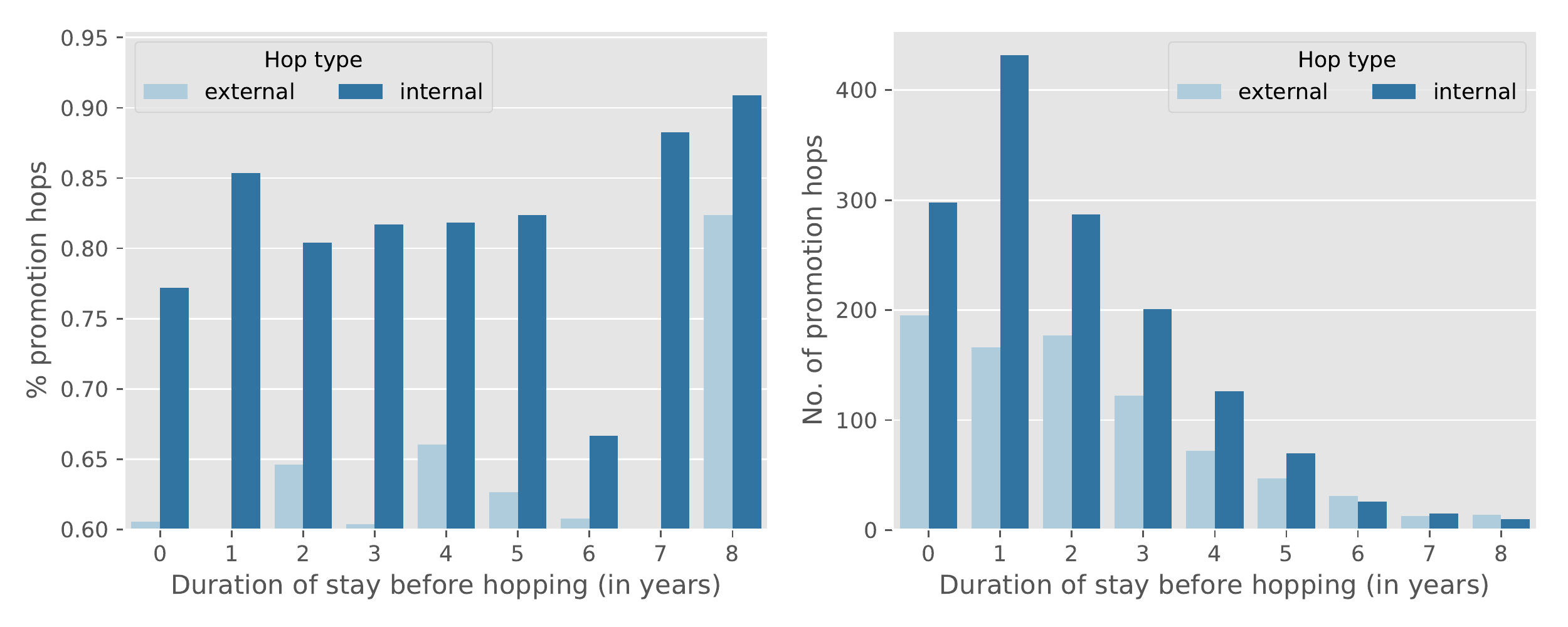} 
}\vfill
\subfloat[Hong Kong]{
  \includegraphics[width=0.9\columnwidth]{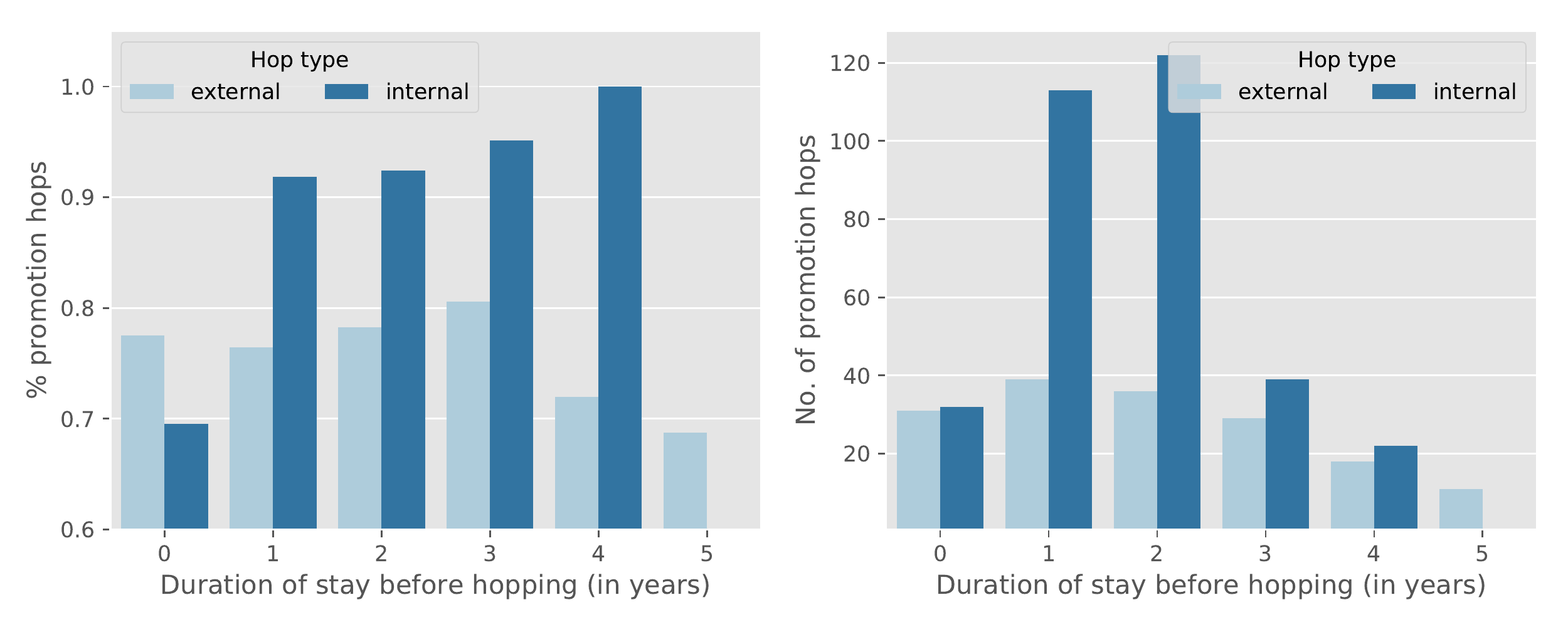} 
}
\caption{Promotion hop fraction and counts for different durations of stay}
\label{fig:hop_frac_vs_duration}
\end{figure} 

Fig.~\ref{fig:hop_level_diff} shows a more fine-grained detail in terms of the level gain distribution. It is evident that the majority of the level gain values are positive, again suggesting that hopping most likely involves promotion rather than demotion (i.e., p(promotion) $>$ p(demotion)). We also found in all datasets that promotions generally give people about 2 years level gain, and demotions generally give people about one year job level loss. The distribution curve at the right charts of figure~\ref{fig:hop_level_diff} shows that the curves for external hops in Singapore and Hong Kong are at the right of that for internal hops. This pattern indicates that external hops in Singapore and Hong Kong tend to give people higher level gain than internal hops. The pattern is quite the opposite for Switzerland. Internal hops in Switzerland results give higher level gain compared to external hops. This finding is consistent with the previous finding in Table~\ref{tab:ch_hop_type_stats}.  

In addition, we investigate whether promotion hops vary with the duration of stay (at some job) before hopping. Fig.~\ref{fig:hop_frac_vs_duration} shows the promotion hop fractions (i.e., p(promotion|external hop) and p(promotion|internal hop)) as well as promotion hop counts as a function of duration of stay prior to hopping. For these plots, we also set the minimum support threshold to filter out unreliable statistics. The right chart of Fig.~\ref{fig:hop_frac_vs_duration} suggests that promotion hops most commonly happen after a person works for 1-2 years. However, the left chart of Fig.~\ref{fig:hop_frac_vs_duration} indicates no obvious relationship between the duration of stay and promotion hop fraction. Regardless, it is again evident that the probability of promotion is higher for internal hops than for external hops.

As with the hop analysis in Section \ref{sec:hop_analysis}, it is possible to further extend the above-mentioned job attribute analysis to company and industry levels. However, we again observe data sparsity issue whereby many small companies and industries in our data have low support, potentially yielding unreliable metrics/statistics computation and inaccurate conclusion. Devising a better analytics approach to deal with data sparsity issue will be left for future work.

\begin{table*}[!t]
\caption{Network statistics of the Singapore dataset}
\label{tab:sg_network_stats}
\begin{tabular}{|l|c|c|}
\hline
Metric       & Job graph & Company graph \\
\hline
\multicolumn{3}{|l|}{Basic} \\
\hline
Number of nodes & 30,531 & 16,112 \\
Number of edges & 45,412 & 43,499 \\
Sparsity of adjacency matrix     & 0.005\% & 0.017\%\\
\hline
\multicolumn{3}{|l|}{Strongly-Connected Component (SCC)} \\
\hline
Number of SCCs                  & 25,340 		  & 11,698\\
Size of the largest SCC         & 5,028 (16.47\%) & 4,367 (27.10\%)\\
Size of the 2nd-largest SCC  	& 11 (0.036\%)     & 3 (0.018\%)\\
\hline
\multicolumn{3}{|l|}{Weakly-Connected Component (WCC)} \\
\hline
Number of WCCs                  & 3,864  		   & 2,187\\ 
Size of the largest WCC        	& 22,201 (72.72\%) & 12,705 (78.85\%)\\
Size of the second-largest WCC  & 22 (0.072\%)      & 6 (0.037\%) \\
\hline 
\end{tabular}
\end{table*}

\begin{table*}[!t]
\caption{Network statistics of the Switzerland dataset}
\label{tab:ch_network_stats}
\begin{tabular}{|l|c|c|}
\hline
Metric       & Job graph & Company graph \\
\hline
\multicolumn{3}{|l|}{Basic} \\
\hline
Number of nodes & 5,867 & 5,704 \\
Number of edges & 8,993 & 10,447 \\
Sparsity of adjacency matrix     & 0.026\% & 0.032\%\\
\hline
\multicolumn{3}{|l|}{Strongly-Connected Component (SCC)} \\
\hline
Number of SCCs                  & 5,139 		& 4,921\\
Size of the largest SCC         & 645 (10.99\%) & 765 (13.41\%)\\
Size of the 2nd-largest SCC  	& 16 (0.27\%)   & 2 (0.035\%)\\
\hline
\multicolumn{3}{|l|}{Weakly-Connected Component (WCC)} \\
\hline
Number of WCCs                  & 1,734  		  & 2,296\\ 
Size of the largest WCC        	& 2,905 (49.51\%) & 2,837 (49.74\%)\\
Size of the second-largest WCC  & 19 (0.32\%)     & 5 (0.088\%) \\
\hline 
\end{tabular}
\end{table*}

\begin{table*}[!t]
\caption{Network statistics of the Hong Kong dataset}
\label{tab:hk_network_stats}
\begin{tabular}{|l|c|c|}
\hline
Metric       & Job graph & Company graph \\
\hline
\multicolumn{3}{|l|}{Basic} \\
\hline
Number of nodes & 1,895 & 1,935 \\
Number of edges & 2,492 & 2,982 \\
Sparsity of adjacency matrix     & 0.069\% & 0.08\%\\
\hline
\multicolumn{3}{|l|}{Strongly-Connected Component (SCC)} \\
\hline
Number of SCCs                  & 1,710 		& 1,762\\
Size of the largest SCC         & 151 (7.97\%)  & 139 (7.18\%)\\
Size of the 2nd-largest SCC  	& 7 (0.37\%)   & 14 (0.72\%)\\
\hline
\multicolumn{3}{|l|}{Weakly-Connected Component (WCC)} \\
\hline
Number of WCCs                  & 484  		    & 892\\ 
Size of the largest WCC        	& 785 (41.42\%) & 763 (39.43\%)\\
Size of the second-largest WCC  & 32 (1.69\%)   & 5 (0.26\%) \\
\hline 
\end{tabular}
\end{table*}

\subsection{Connectivity Analysis}

\textbf{Network structure analysis.} In this section, we analyze the job hop behavior at the network level, which includes job and organization graphs. We set edge minimum support equals to two for this analysis. The basic statistics of the job and organization graphs are summarized in Table~\ref{tab:sg_network_stats},~\ref{tab:ch_network_stats}, and~\ref{tab:hk_network_stats}. We can conclude that all talent flow network graphs are sparse in general, having small number of edges relative to the squared number of nodes. We also examine the connectedness of the graphs by looking at the strongly-connected component (SCC) and weakly-connected component (WCC) metrics. The former checks for connectedness by following the directionality of the graph edges, whereas the latter ignores the directionality.

Overall, the results in Table~\ref{tab:sg_network_stats},~\ref{tab:ch_network_stats}, and~\ref{tab:hk_network_stats} indicate that there exists a giant component for both job and organization graphs, and its size is significantly bigger than the second largest component. As such, we can conclude that our job and organization graphs are fairly well-connected, in the sense that there exists a path between any two nodes within the giant components.

\begin{figure}[!t]
\centering
\subfloat[Singapore]{
  \includegraphics[width=0.9\columnwidth]{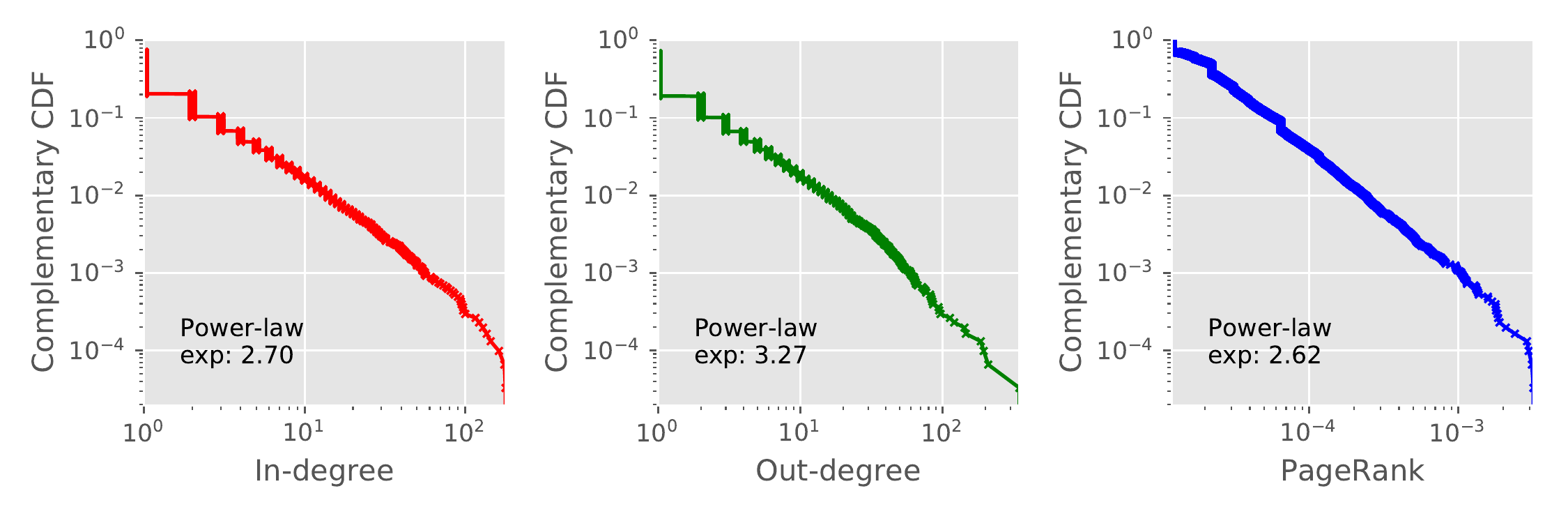} 
}\vfill
\subfloat[Switzerland]{
  \includegraphics[width=0.9\columnwidth]{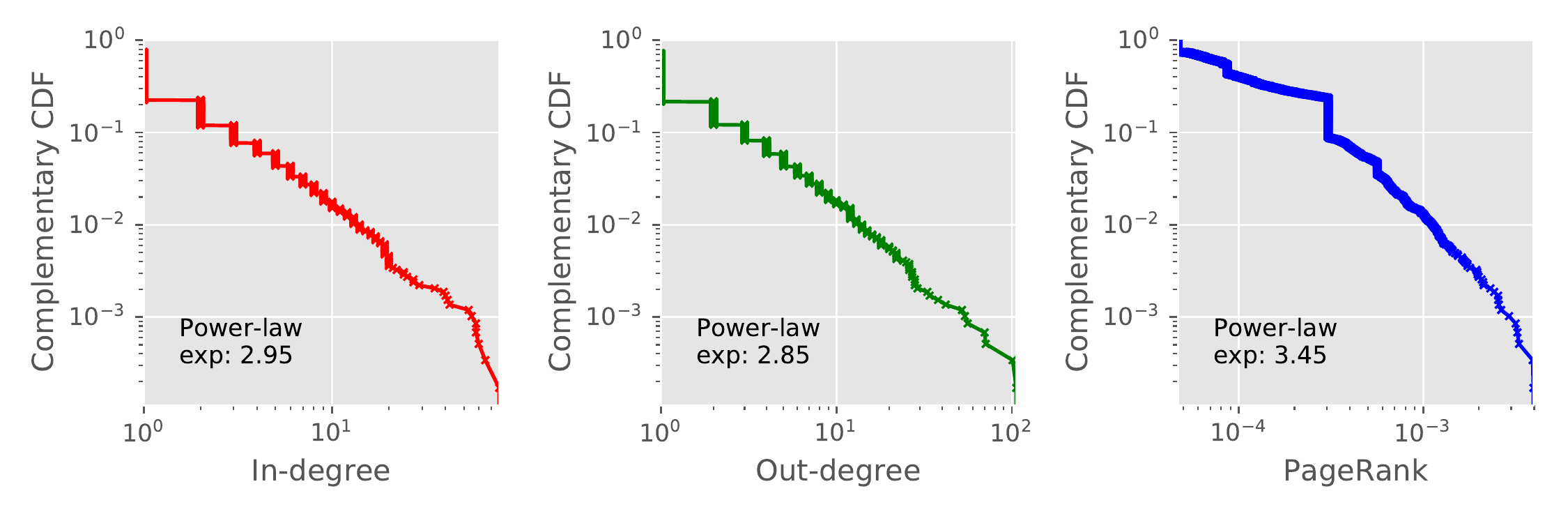} 
}\vfill
\subfloat[Hong Kong]{
  \includegraphics[width=0.9\columnwidth]{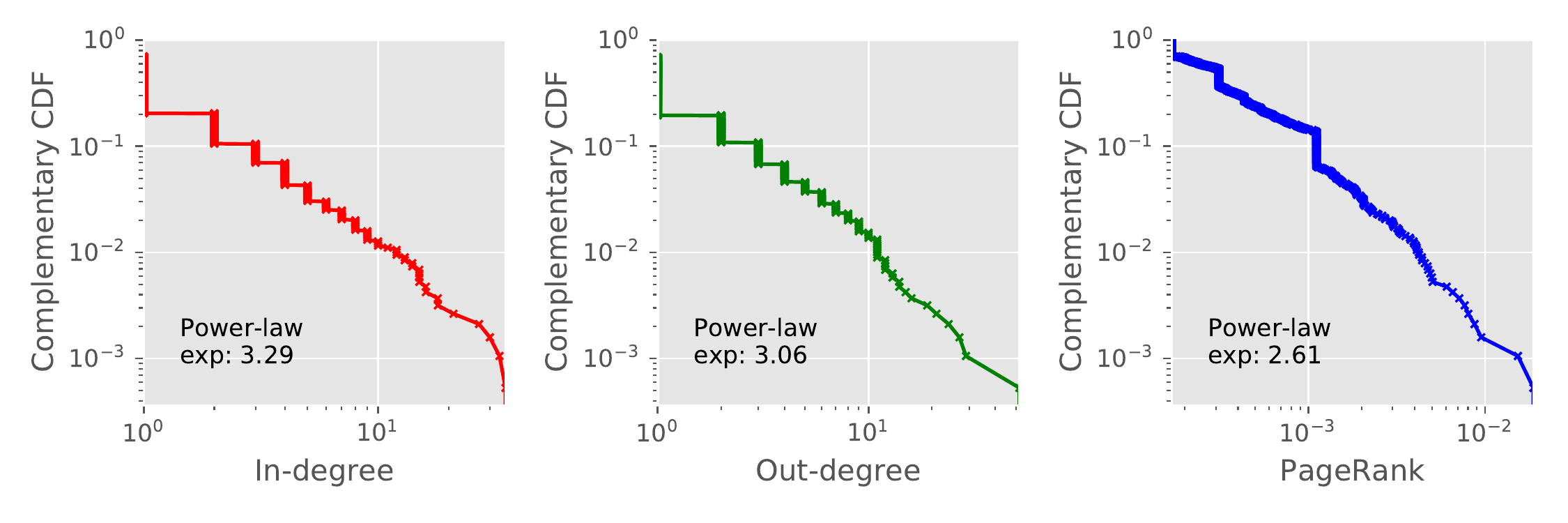} 
}
\caption{Centrality distribution of job hop graph}
\label{fig:job_hop_centrality_dist}
\end{figure}

With the connectedness trait validated, we now examine the centrality properties of the nodes in our hop graphs. Fig.~\ref{fig:job_hop_centrality_dist} presents the complementary cumulative distribution functions (CDFs) of the in-degree, out-degree, and PageRank centralities for the job graph. It is shown that all three metrics exhibit heavy-tail, skewed distribution. We performed power-law fitting and obtained exponent terms of greater than 2 for all graphs, thereby indicating a scale-free phenomenon. Similar result was obtained for the organization graph, although the results are not shown here due to space constraint.

\begin{figure}[!t]
\centering
\subfloat[Top jobs in Singapore \label{fig:sg_job_centrality_top}]{
  \includegraphics[width=0.48\columnwidth]{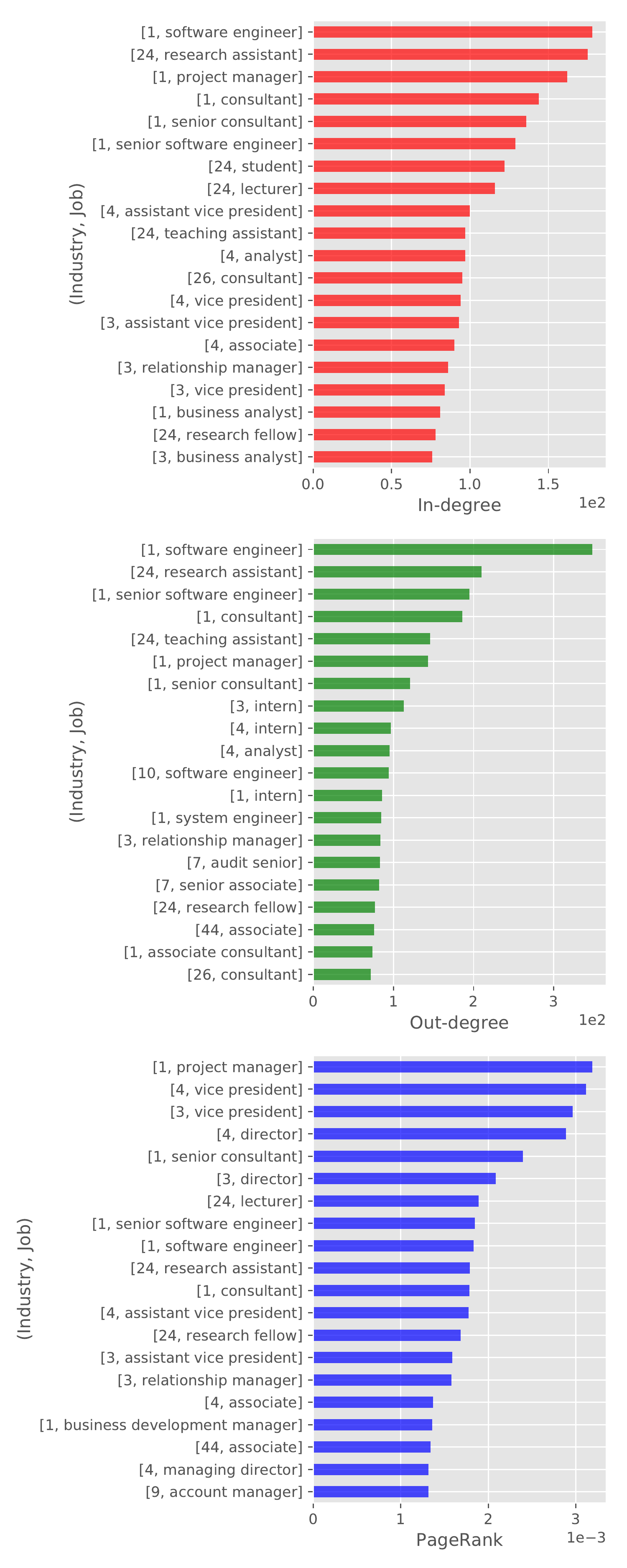}
}\hfill
\subfloat[Top companies in Singapore \label{fig:sg_com_centrality_top}]{
  \includegraphics[width=0.48\columnwidth]{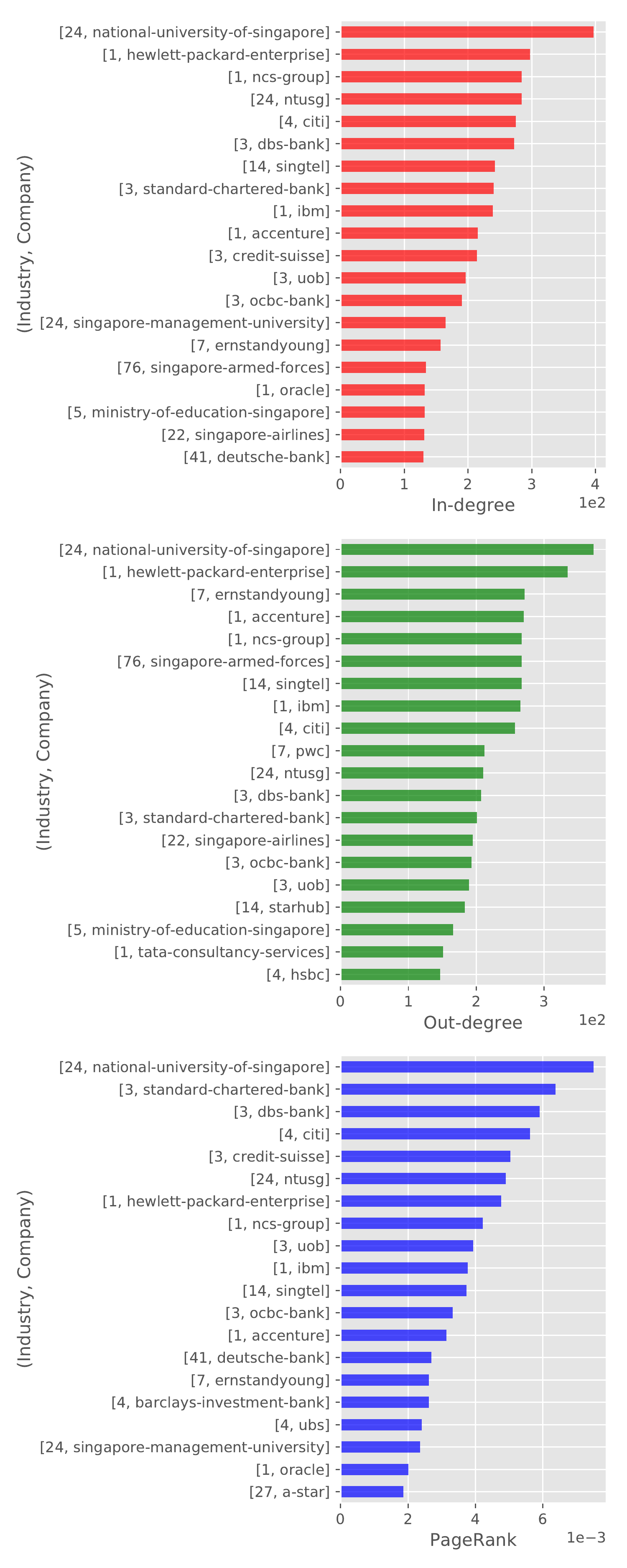} 
}
\caption{Centrality of top jobs and companies in Singapore}
\label{fig:sg_hop_centrality_dist}
\end{figure}

\begin{figure}[!t]
\centering
\subfloat[Top jobs in Switzerland\label{fig:ch_job_centrality_top}]{
  \includegraphics[width=0.48\columnwidth]{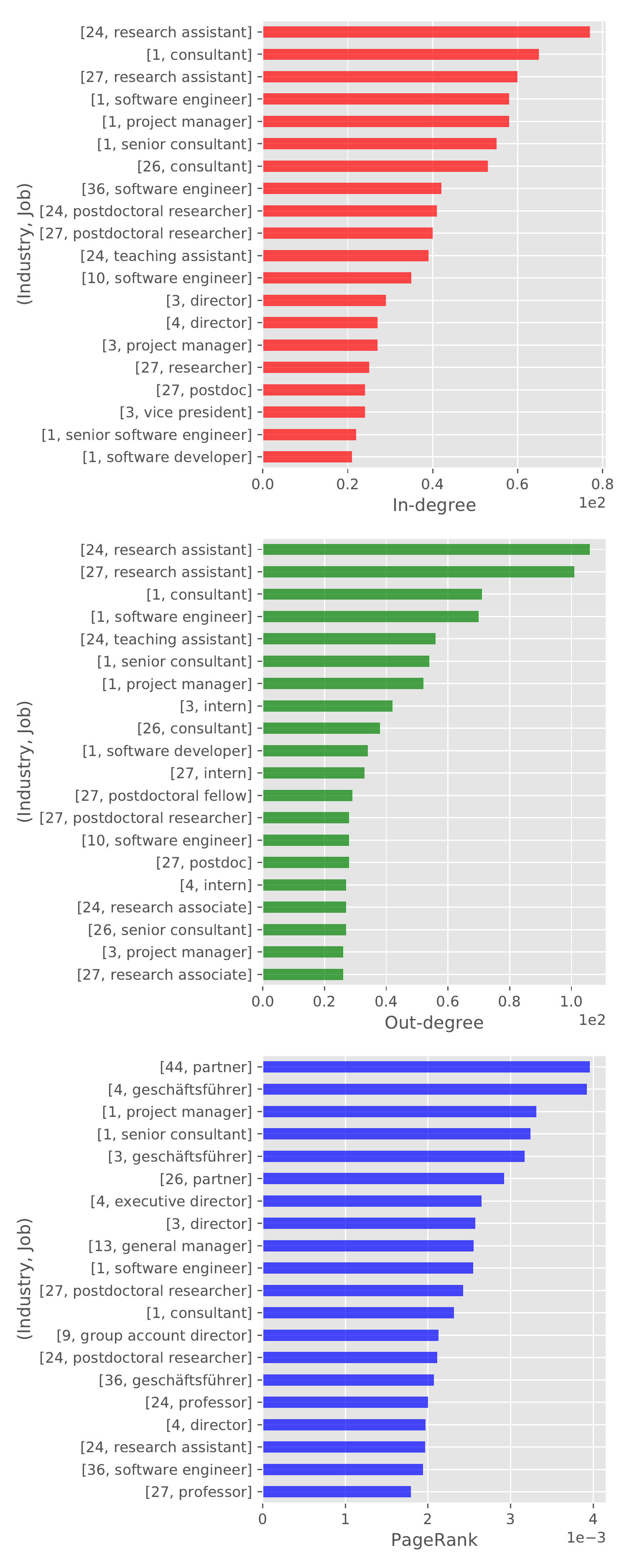}
}\hfill
\subfloat[Top companies in Switzerland\label{fig:ch_com_centrality_top}]{
  \includegraphics[width=0.48\columnwidth]{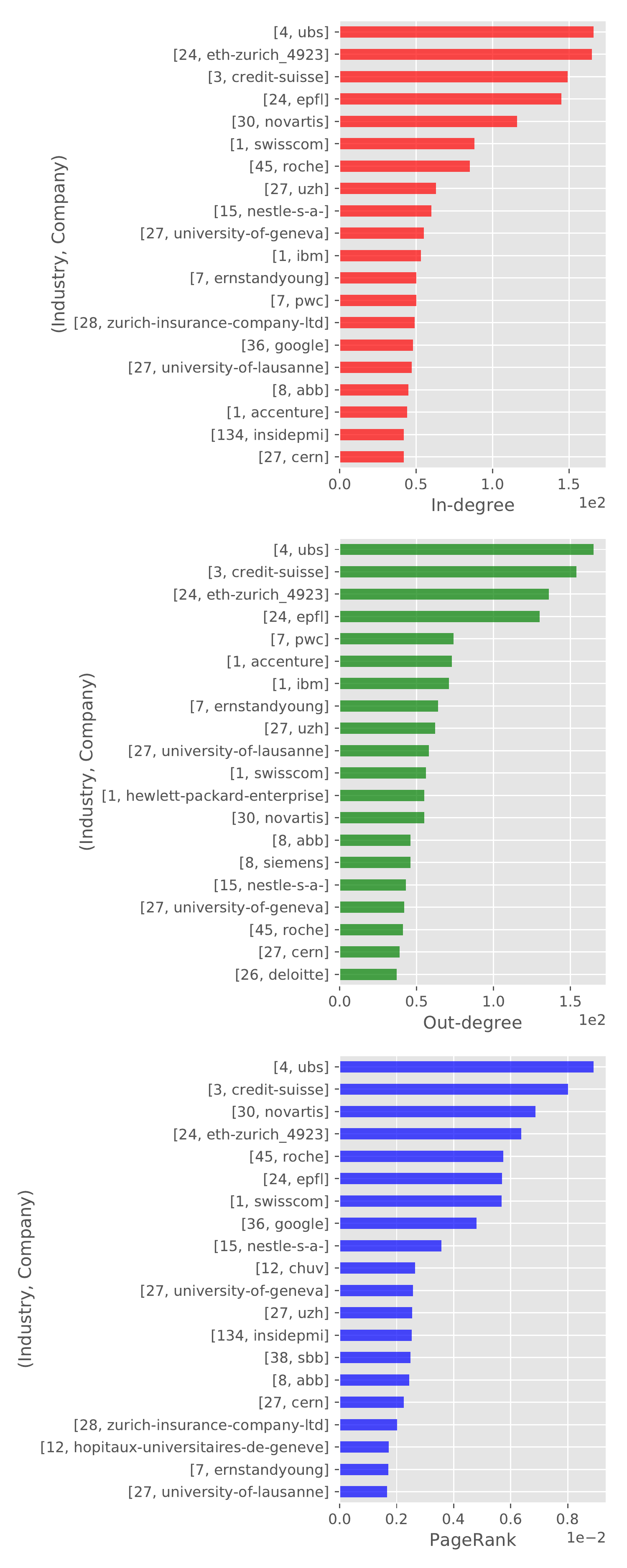} 
}
\caption{Centrality of top jobs and companies in Switzerland}
\label{fig:ch_hop_centrality_dist}
\end{figure}

\begin{figure}[ht]
\centering
\subfloat[Top jobs in Hong Kong \label{fig:hk_job_centrality_top}]{
  \includegraphics[width=0.48\columnwidth]{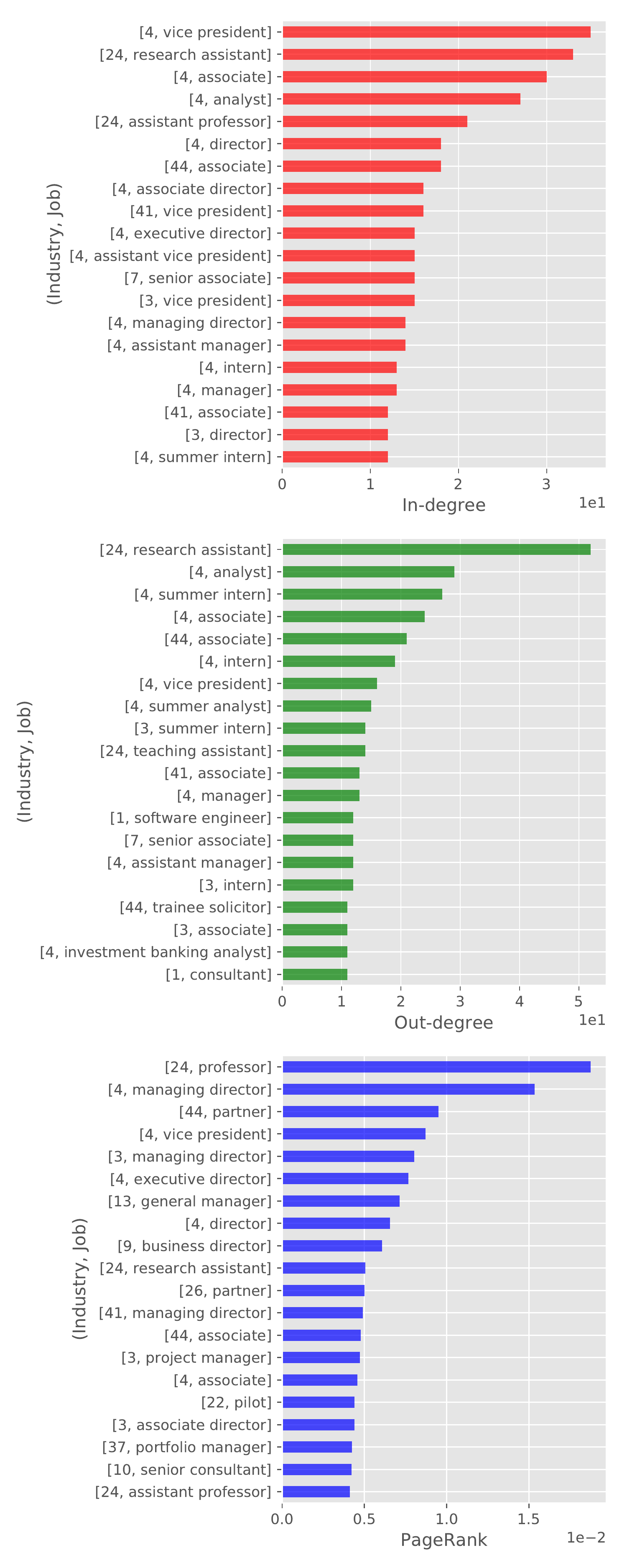}
}\hfill
\subfloat[Top companies in Hong Kong \label{fig:hk_com_centrality_top}]{
  \includegraphics[width=0.48\columnwidth]{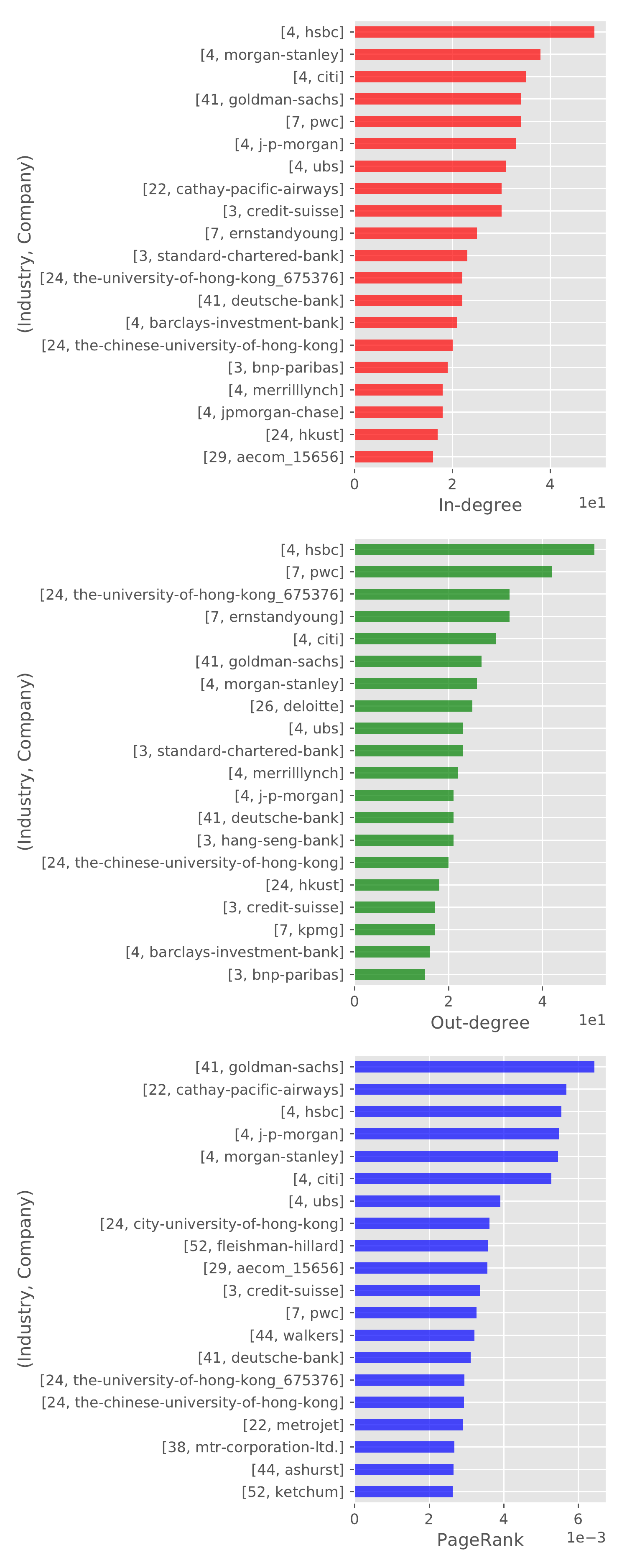} 
}
\caption{Centrality of top jobs and companies in in Hong Kong}
\label{fig:hk_hop_centrality_dist}
\end{figure}

\textbf{Job centrality analysis}. Next, we evaluate the top nodes having the highest centrality values in the job-level and organization-level graphs for Singapore, Switzerland, and Hong Kong, as shown in Figs.~\ref{fig:sg_hop_centrality_dist}, ~\ref{fig:ch_hop_centrality_dist}, and~\ref{fig:hk_hop_centrality_dist} respectively. The results provide several interesting insights. For the job graph, we find that the top in-degree, out-degree and PageRank jobs are overall dominated by major industries\footnote{Major industry codes: 1 = Information Technology and Services, 3 = Banking, 4 = Financial Services, 7 = Accounting, 10 = Computer Software, 24 = Higher Education, 26 = Management Consulting}.

From the left charts of Fig. \ref{fig:sg_hop_centrality_dist}(a), \ref{fig:ch_hop_centrality_dist}(a) and \ref{fig:hk_hop_centrality_dist}(a), we can see that the top in-degree nodes refer to those popular jobs in major industries that attract talents. Meanwhile, the middle charts of Fig. \ref{fig:sg_hop_centrality_dist}(a), \ref{fig:ch_hop_centrality_dist}(a) and \ref{fig:hk_hop_centrality_dist}(a) suggest that the top out-degree jobs are those that involve versatile skills (e.g., software engineer, consultant) or interim roles (e.g., intern). People having these jobs may thus be able to move to more diverse range of jobs/organizations (i.e., talent supplier). Finally, the right charts of Fig. \ref{fig:sg_hop_centrality_dist}(a), \ref{fig:ch_hop_centrality_dist}(a) and \ref{fig:hk_hop_centrality_dist}(a) show that the top PageRank nodes correspond to high-level, managerial jobs (e.g., Director, Manager, Vice President). This conforms with our intuition on PageRank as a measure of job desirability (cf. Section \ref{sub:network-analysis}).

\textbf{Organization centrality analysis.} Fig. \ref{fig:sg_hop_centrality_dist}(b), \ref{fig:ch_hop_centrality_dist}(b) and \ref{fig:hk_hop_centrality_dist}(b) show the top companies in the three countries/regions based on in-degree, out-degree and PageRank. The top companies returned by these measures are large corporations.  Given that different set of companies operate in these countries/regions, it is not feasible to compare top companies across countries/regions. It is however noted that the top few companies of each country can be quite different when applying the different measures.  Among the three countries/regions, Switzerland seems to have more top company overlaps between the measures.  

%% file: conclusion.tex
 \section{Conclusion}
\label{sec:conclusion}

In this paper, we put forward a data analytics approach to study job hops at a large scale using the OPN data from multiple countries/regions. In conclusion, our study provides a few key takeaways: 

\begin{itemize}
\item We discover that external hops are not necessarily influenced by work experience, job age, and number of skills. We also observe that external hops are very common, and the Singapore workforce exhibits the highest external hop fraction among all the three countries/regions studied in this work. 

\item Our analysis on hop classification and job attribute demonstrate that: (1) external hops are very common; (2) Job hopping involve promotions more likely than demotions, and people are more likely to get promoted due to internal hops than getting promoted due to external hops; (3) promotion hops most commonly happen after a person works for 1-2 years.

\item From our network connectivity analyses, we find that: (1) top in-degree job (organization) nodes are prominent jobs (companies) that attract talents, whereas top out-degree job (organization) nodes are influential jobs (organizations) that  supply talents; and (2) job (organization) nodes with high PageRank refer to desirable, major jobs (organizations) that are well-known for providing good career offering.
\end{itemize}

Our comparative study on the OPN data from Singapore, Switzerland and Hong Kong has also enabled us to gain additional insights on the unique characteristics of the workforces in different countries/regions, such as:
\begin{itemize}
\item For the same job title, jobs in Switzerland tends to have 2-3 longer years of experiences than jobs in Singapore, whereas jobs in Hong Kong tend to have more or less comparable years of experience to jobs in Singapore.
\item Most jobs in Switzerland tend to have longer job ages than those in Singapore users, suggesting that, for the same job title, the Switzerland workforce is ahead of the Singapore counterparts in terms of job establishment. In contrast, the Hong Kong users tend to have lower job age than the Singapore users, implying that jobs in Hong Kong are generally less established than those in Singapore.
\item The resulting statistics of external and internal hops suggest that the Switzerland and Hong Kong workforces are generally more ``loyal'' than the Singapore workforce. This is evident from the significantly higher proportion of external hops (relative to internal hops) in the Singapore data.
\end{itemize}

The findings from this paper lead to a few possibilities.  Firstly, we demonstrate that it is possible to re-purpose the career histories of OPN profiles to study the job hop patterns of workforce within a country/region, and to  compare across countries/regions. This vastly improves the scale and granularity of job hop study, which was traditionally done using surveys. Through our analysis, we show that the propensity to perform job hops is relatively higher among the young workforce than the older one.  This could lead to two main concerns, namely: (i) the limited time to acquire adequate skills on the job among the young employees; and (ii) the unwillingness of companies to provide them skill training.  These concerns may cost the workforce long-term's skill development and productivity. To overcome these, more incentives may be introduced to encourage young employees to stay longer on their jobs.  One could also increase the chance of job promotions among the younger employees.

Finally, our analysis also shows that job and organization graphs are well connected. We further define job centrality measures to determine attractive jobs and companies.  Such measures allow jobs and companies to be ranked for applicants' reference during job search.  These measures can also be further refined to find attractive jobs and companies in specific industry domains.